\documentclass[11pt]{article}
\usepackage{amssymb}
\usepackage{graphics}
\usepackage{epsfig}
\usepackage{a4wide}
\usepackage{slashed}

\begin{document}

\newcommand{\ppvq}{$pp \to W_H(Z_H) q_- + X$ }
\newcommand{\ppww}{$pp \to W_H^+ W_H^- + X$ }
\newcommand{\qqww}{$q\bar q \to W_H^+ W_H^-$}
\newcommand{\uuww}{$u\bar u \to W_H^+ W_H^-$}
\newcommand{\ggww}{$gg \to W_H^+ W_H^-~$}
\newcommand{\qqwwg}{$q\bar q \to W_H^+ W_H^- + g$}
\newcommand{\qgwwq}{$q(\bar q)g \to W_H^+ W_H^- + q(\bar q)$}

\title{$W_H$-pair Production in the Littlest Higgs Model with T parity in next-to-leading order QCD at LHC  }
\author{ Du Song-Ming, Guo Lei, Liu Wen, Ma Wen-Gan, and Zhang Ren-You \\
{\small  Department of Modern Physics, University of Science and Technology}  \\
{\small  of China (USTC), Hefei, Anhui 230026, P.R.China}}

\date{}
\maketitle \vskip 15mm
\begin{abstract}
In the framework of the littlest Higgs model with $T$ parity, we
study the $W_H$-pair production at the CERN Large Hadron Collider up
to the QCD next-to-leading order (NLO). The kinematic distributions
of final decay products and the theoretical dependence of the cross
section on the factorization/renormalization scale are analyzed. We
adopt the PROSPINO scheme in the QCD NLO calculations to avoid
double counting and keep the convergence of the perturbative QCD
description. Our numerical results show that the QCD NLO corrections
significantly reduce the scale uncertainty, and enhance the leading
order integrated cross section with a $K$-factor in the range of
$1.10-1.22$ ($1.09-1.17$) with the symmetry breaking scale $f$
varying from $400~GeV$ ($400~GeV$) to $1.5~TeV$ ($1.0~TeV$) at the
$14~TeV$ $(8~TeV)$ LHC. We find that it is possible to select the
signal events of the $W_H$-pair production from the $pp\to W^+ W^-
\to e^+ \mu^-\nu_{e}\bar{\nu}_{\mu}+X$ background with high ratio of
signature over background by taking proper lower limits on
transverse momenta, invariant mass of the final charged leptons and
the missing transverse momentum.
\end{abstract}


\vfill \eject \baselineskip=0.32in

\renewcommand{\theequation}{\arabic{section}.\arabic{equation}}
\renewcommand{\thesection}{\Roman{section}.}
\newcommand{\nb}{\nonumber}

\newcommand{\Dir}{\kern -6.4pt\Big{/}}
\newcommand{\Dirin}{\kern -10.4pt\Big{/}\kern 4.4pt}
\newcommand{\DDir}{\kern -7.6pt\Big{/}}
\newcommand{\DGir}{\kern -6.0pt\Big{/}}

\makeatletter      
\@addtoreset{equation}{section}
\makeatother       

\par
\section{Introduction}
\par
Although the standard model (SM) \cite{s1,s2} provides a remarkably
successful description of high energy physics phenomena at the
energy scale up to $100~GeV$, it leaves a number of theoretical
problems unsolved. Many extended models are proposed to deal with
these problems such as grand unified theories \cite{model-1},
supersymmetric models \cite{model-2}, extra dimensions models
\cite{model-3}, left-right symmetric models \cite{model-5}, B-L (baryon
number minus lepton number)
extended SM models \cite{model-6}, little Higgs models
\cite{LittleHiggs} and many more. Each of these models has
motivation to solve one or more of the problems that the SM
encounters. Among them the little Higgs models deserve attention due
to their elegant solution to hierarchy problem and are proposed as
one kind of electroweak symmetry breaking (EWSB) models without
fine-tuning in which the Higgs boson is naturally light as a result
of nonlinearly realized symmetry \cite{Arkani}-\cite{LH7}. The
littlest Higgs (LH) model \cite{LH4}, an $SU(5)/SO(5)$ nonlinear
sigma model \cite{LH5}, is the most simplest version of little Higgs
models, in which a set of new heavy gauge bosons $(A_H,W_H,Z_H)$ and
a vector-like quark $(T)$ are introduced to cancel the quadratic
divergence contribution to Higgs boson mass from the SM gauge boson
loops and the top quark loop respectively. However, this model
predicts large corrections to electroweak precision observables and
the scale of the global symmetry breaking $f$, is constrained by
experimental data \cite{Limit-f}, which set severe constraints on
the new heavy particle masses and the model parameters. For
instance, recent experimental measurements on the decay processes of
$W_H^{\mp}\to l^{\mp}\stackrel{(-)}{\nu}$ and $Z_H\to l^{+}l^{-}$
provide the constraints of $M_{W_H}> 2.18~TeV$ and $M_{Z_H}>
1.83~TeV$ \cite{Wmass,Zmass}. These constraints would enforce the
symmetry breaking scale $f$, which characterizes the mass of new
particles, to be larger than $2.5~TeV$ and $3~TeV$ respectively.
Consequently, the cutoff scale $\Lambda \sim 4\pi f$ becomes so
large that calls for the fine-tuning between the electroweak scale
and the cutoff scale again.

\par
By introducing a discrete symmetry, the $T$ parity, the littlest
Higgs model with $T$ parity (LHT)
\cite{Low:2004xc}-\cite{Cheng:2003ju} offers a viable solution to
the naturalness problem of the SM, and also predicts a set of new
heavy fermions, gauge bosons as well as a candidate for dark matter.
In the LHT, all the SM particles are $T$-even and almost all the new
heavy particles are $T$-odd. Due to the different $T$ parity quantum
numbers, the SM gauge bosons cannot mix with the new gauge bosons in
the LHT. This would alleviate the constraints from the electroweak
precision tests and thus allows the scale $f$ to be significantly
lower than $1~TeV$ \cite{Hubisz:2005tx}. For instance, due to the
$T$ parity conservation, the processes $W_H^{\mp}\to
l^{\mp}\stackrel{(-)}{\nu}$ and $Z_H\to l^{+}l^{-}$ are forbidden,
and the only decay modes of these $T$-odd heavy gauge bosons are
$W_H\to A_H W$ and $Z_H\to A_H H$. In this case, the leptons are
produced from the decays of $W$ and $H$, but not from the heavy
gauge bosons directly. Therefore, these $T$-even gauge bosons escape
from the experimental constraints shown in Refs.\cite{Wmass,Zmass}.
Furthermore, as a lightest $T$-odd particle, the heavy photon $A_H$
cannot further decay into other particles, and would be a good
candidate for the dark matter \cite{Asano}. Since the CERN Large
Hadron Collider (LHC) has potential to detect the signals of new
gauge bosons and fermions, the phenomenology of the LHT would be
quite interesting and a number of phenomenological works has been
presented
\cite{Hubisz:2004ft,cpyuan:2006ph,sasha_pheno,Chen:2006ie}.
Recently, the QCD NLO corrections to the process \ppvq has been
presented in Ref.\cite{YanH}. Of all heavy gauge boson production
processes, the heavy gauge boson $W_H$-pair production can be
particularly significant due to the potential of its copious
productions at the LHC as shown in Refs.\cite{cpyuan:2006ph,qhcao},
where the $W_H$-pair production at the LHC is studied at the
leading-order (LO).

\par
In this paper, we make a precision investigation for the process
\ppww at the LHC including the QCD NLO corrections. In Sec.II we
make a brief review of the relevant theory of the LHT. The detailed
strategies of the calculation are given in Sec.III. The numerical
results and discussions are presented in Sec.IV. Finally we present
a short summary.

\vskip 5mm
\section{The related LHT theory }\label{theory}
\par
Before our calculations, we will briefly recapitulate the LHT theory
which is relevant to the analysis in this work. The details of
the LHT can be found in
Refs.\cite{Low:2004xc,Hubisz:2004ft,Hubisz:2005tx,cpyuan:2006ph}.

\par
At some high scale $f$ the global symmetry $SU(5)$ is broken down to
$SO(5)$, leading to 14 massless Nambu-Goldstone bosons. Four of them
are manifested as the longitudinal modes of the heavy gauge bosons.
The other 10 decompose into a $T$-even $SU(2)$ doublet $h$, identified
as the SM Higgs field, and a complex $T$-odd $SU(2)$ triplet
$\Phi$, which obtains a mass of $m_{\Phi} = \sqrt{2}m_hf/v_{SM}$,
with $m_h$ and $v_{SM}$ being SM Higgs mass and the electroweak
symmetry break scale, respectively.

\par
The additional discrete symmetry, $T$-parity, is in analogy to the
$R$-parity in the minimal supersymmetric standard model (MSSM)
\cite{Low:2004xc,Hubisz:2004ft,Cheng:2003ju}. The $T$-parity
transformations for gauge sector are defined as the exchange between
the gauge bosons of the two $SU(2)\times U(1)$ groups, i.e., $W_1^a
\leftrightarrow W_2^a$ and $B_1 \leftrightarrow B_2$. Thus their
$T$-odd and $T$-even combinations can be obtained as
\begin{eqnarray}
W_H^a=\frac{1}{\sqrt{2}}(W_1^a-W_2^a),&
B_H=\frac{1}{\sqrt{2}}(B_1-B_2),& (T-odd), \nb\\
W_L^a=\frac{1}{\sqrt{2}}(W_1^a+W_2^a),&
B_L=\frac{1}{\sqrt{2}}(B_1+B_2),& (T-even).
\end{eqnarray}
The mass eigenstates of the gauge sector in the LHT are expressed as
\begin{eqnarray}
W_H^{\pm}=\frac{1}{\sqrt 2}(W_H^1\mp i W_H^2),& Z_H=s_H B_H+c_H
W^3_H,&
A_H=c_H B_H-s_H W^3_H , \nb \\
W_L^{\pm}=\frac{1}{\sqrt 2}(W_L^1\mp i W_L^2),& Z_L=-s_w B_L+c_w
W^3_L,& A_L=c_w B_L+s_w W^3_L,
\end{eqnarray}
where $s_{w}=\sin{\theta_{W}}$, $c_{w}=\cos{\theta_{W}}$,
$s_{H}=\sin{\theta_{H}}$, $c_{H}=\cos{\theta_{H}}$, $\theta_W$ is
the Weinberg angle, and the mixing angle $\theta_H$ at the ${\cal
O}(v^2/f^2)$ is expressed as
\begin{eqnarray}
\sin \theta_H \simeq \left[ \frac{5gg'}{4(5g^2-g'^2)}\frac{v_{SM}
^2}{f^2} \right].
\end{eqnarray}
Then the gauge sector consists of $T$-odd heavy new gauge bosons
$W_H^{\pm}$, $Z_H$, $A_H$ and $T$-even light gauge bosons identified
as SM gauge bosons, $W^{\pm}$, $Z^0$ and one massless photon. The
$T$ parity partner of the photon, $A_H$, is the lightest $T$-odd
particle, therefore, the candidate of dark matter in the LHT. The
masses of the $T$ parity partners of the photon, $Z^0$- and
$W^{\pm}$-boson are expressed as \cite{cpyuan:2006ph}
\begin{eqnarray} \label{m_v}
m_{W_H}\simeq m_{Z_H} \simeq
gf\left(1-\frac{1}{8}\frac{v_{SM}^2}{f^2}\right),& m_{A_H}\simeq
\frac{1}{\sqrt{5}}g'f\left(1-\frac{5}{8}\frac{v_{SM}^2}{f^2}\right),
\end{eqnarray}
where $v_{SM}= 246~GeV$. At the tree level the SM gauge boson masses
can be expressed as $m_W=\frac{gv_{SM}}{2}$ and
$m_Z=\frac{v_{SM}\sqrt{g^2+g^{\prime 2}}}{2}$.

\par
In the LHT, the fermion sector of the first two generations in the
SM is remained unchanged and the third generation of quarks is
modified. We introduce two fermion doublets $q_1$ and $q_2$ for each
fermion generation. The $T$ parity transformation to these fermion
doublets is defined as $q_1 \leftrightarrow  - q_2$. Therefore, the
$T$-odd and $T$-even combinations can be constructed as
$q_-=\frac{1}{\sqrt{2}}(q_1+q_2)$ and
$q_+=\frac{1}{\sqrt{2}}(q_1-q_2)$, where $q_+$ is the doublet for
the SM fermions and $q_-$ for their $T$-odd partners. We take the
Lagrangian suggested in
Refs.\cite{Low:2004xc,Hubisz:2004ft,Hubisz:2005tx} to generate the
masses of the $T$-odd fermion doublets,
\begin{eqnarray}
 -\kappa f (\bar{\Psi}_2 \xi \Psi_c
+\bar{\Psi}_1 \Sigma_0 \Omega \xi^\dagger \Omega\Psi_c)+{\rm h.c.},
\label{Lagrangian}
\end{eqnarray}
where $\Omega = diag(1,1,-1,1,1)$,
$\Psi_c=(q_c,\chi_c,\tilde{q}_c)^T$, and the $SU(5)$ multiplets
$\Psi_1$ and $\Psi_2$ are expressed as
\begin{equation}\label{Psi}
\begin{array}{ccc}
{\Psi}_1=\left( \begin{array}{c} q_1 \\ 0 \\ {\bf 0}_2
\end{array}\right) \,,& {\Psi}_2=\left(\begin{array}{c} {\bf 0}_2 \\
0 \\ q_2
\end{array}\right).
\end{array}
\end{equation}
The interaction Lagrangian in Eq.(\ref{Lagrangian}) can be proofed
to be invariant under $T$-parity, and $T$-odd quark doublet $q_-$
gets a Dirac mass with $\tilde{q}_c\equiv(id_{R_-},-iu_{R_-})^{\rm
T}$ from Eq.(\ref{Lagrangian}) expressed as \cite{cpyuan:2006ph}
\begin{eqnarray} \label{m_q}
m_{U_-}\simeq \sqrt{2}\kappa
f\left(1-\frac{1}{8}\frac{v_{SM}^2}{f^2}\right), &&
m_{D_-}=\sqrt{2}\kappa f,
\end{eqnarray}
where the lower indexes $U_-=u_-,c_-,t_-$ and $D_-=d_-,s_-,b_-$,
which represent the $T$-odd heavy partners of the SM quarks, and
$\kappa$ is the mass coefficient in Lagrangian of the quark sector.
As we know in the LHT $f > 500~GeV$ \cite{Hubisz}, it is evident
from Eq.(\ref{m_q}) that the $T$-odd up- and down-type heavy
partners have nearly equal masses.

\par
In order to avoid the large radiative correction to Higgs boson mass
induced by top-quark loop, the top sector must be additionally
modified. We introduce the following two multiplets,
\begin{equation}
\begin{array}{ccc}
{\cal Q}_1=\left( \begin{array}{c} q_1 \\ U_{L1} \\ {\bf 0}_2
\end{array}\right) \,,& {\cal Q}_2=\left(\begin{array}{c} {\bf 0}_2 \\
U_{L2} \\ q_2
\end{array}\right),
\end{array}
\end{equation}
where $U_{L1}$ and $U_{L2}$ are the singlet fields and the $q_1$ and
$q_2$ are the doublets. Under the $SU(5)$ and the $T$ parity
transformations, ${\cal Q}_1$ and ${\cal Q}_2$ behave themselves
same as $\Psi_1$ and $\Psi_2$.

\par
In addition to the $T$-even SM top quark right-handed $SU(2)$
singlet $u_R$, the LHT contains two $SU(2)$ singlet fermions
$U_{R1}$ and $U_{R2}$ of hypercharge 2/3, which transform under $T$
parity as
\begin{equation}
U_{R1}\leftrightarrow -U_{R2}.
\end{equation}
The T parity invariant Yukawa Lagrangian of the top sector can be
written as
\begin{eqnarray}\label{top-yukawa}
{\cal L}^{Y}_t &=& \frac{\lambda_1 f }{2\sqrt{2}}\epsilon_{ijk}
\epsilon_{xy} \big[ (\bar{{\cal Q}}_1)_i \Sigma_{jx} \Sigma_{ky}  -
(\bar{{\cal Q}}_2 \Sigma_0)_i \tilde{\Sigma}_{jx}
\tilde{\Sigma}_{ky} \big] u_R \nonumber \\ &&  +
\lambda_2 f (\bar{U}_{L1} U_{R1} + \bar{U}_{L2} U_{R2})+ {\rm
h.c.}~.
\end{eqnarray}
where $\tilde{\Sigma}=\Sigma_0\Omega \Sigma^\dagger \Omega \Sigma_0$
is the image of the $\Sigma$ field under $T$ parity, and $i,~j$ and
$k$ run over $1-3$ and $x$ and $y$ over $4-5$. The $T$ parity
eigenstates are constructed as
\begin{equation}\label{eigenstate}
q_\pm = \frac{1}{\sqrt{2}}(q_1 \mp q_2),~~~~ U_{L\pm} =
\frac{1}{\sqrt{2}}(U_{L1} \mp U_{L2}),~~~~ U_{R\pm} =
\frac{1}{\sqrt{2}}(U_{R1} \mp U_{R2}).
\end{equation}
The $T$-odd states $U_{L-}$ and $U_{R-}$ combine to form a Dirac
fermion $T_-$, and we obtain the mass of the $T_-$ quark from the
Lagrangian of Eq.(\ref{top-yukawa}) as
\begin{equation}\label{T-oddMass}
m_{T_-}=\lambda_2 f.
\end{equation}
The left-handed (right-handed) top quark $t$ is a linear combination
of $u_{L_+}$ and $U_{L_+}$ ($u_{R+}$ and $U_{R_+}$), and another
independent linear combination is a heavy $T$-even partner of the
top quark $T_+$:
\begin{eqnarray}
\left(
\begin{array}{c}
t_X\\
T_{+X}
\end{array}
\right)&=&
\left(
\begin{array}{cc}
c_X & -s_X\\
s_X & c_X
\end{array}
\right) \left(
\begin{array}{c}
u_{X_+}\\
U_{X_+}
\end{array}
\right),~~(X=L,R),
\end{eqnarray}
where the mixing matrix elements are approximately expressed as
\begin{eqnarray}
s_L = s_\alpha^2 \frac{v_{SM}}{f}+\cdots,~~ s_R = s_\alpha\left[
1-\frac{c_\alpha^2(c_\alpha^2-s_\alpha^2)}{2}\frac{v_{SM}^2}{f^2}+\cdots\right].
\label{s-LR}
\end{eqnarray}
There we define $s_\alpha=\lambda_1/\sqrt{\lambda_1^2+\lambda_2^2}$
and $c_\alpha=\lambda_2/\sqrt{\lambda_1^2+\lambda_2^2}$. The $t$ is
identified with the SM top and $T_+$ is its $T$-even heavy partner.
Then the masses of the top quark and $T$-even heavy top quark can be
obtained as
\begin{equation}\label{t&T-evenMass}
m_t\simeq \frac{ \lambda_1 \lambda_2
v_{SM}}{\sqrt{\lambda_1^2+\lambda_2^2}},~~~~m_{T_+} \simeq
f \sqrt{\lambda_1^2+\lambda_2^2}.
\end{equation}

\par
The Feynman rules in the LHT related to our calculations
are presented in Appendix A.

\vskip 5mm
\section{Calculation descriptions }\label{calc}
\par
In this work, we adopt the five-flavor scheme (5FS) in the LO and
QCD NLO calculations and neglect the masses of the $u,d,c,s,b$
quarks. In our calculations we use the 't Hooft-Feynman gauge and
employ developed FeynArts 3.4 package \cite{FeynArts} to generate
Feynman diagrams and their corresponding amplitudes. The reduction
of output amplitudes are implemented by FormCalc-5.4 package
\cite{FormCalc}.

\par
\subsection{LO cross section }
\par
\begin{figure}
\begin{center}
\includegraphics[width=0.6\textwidth]{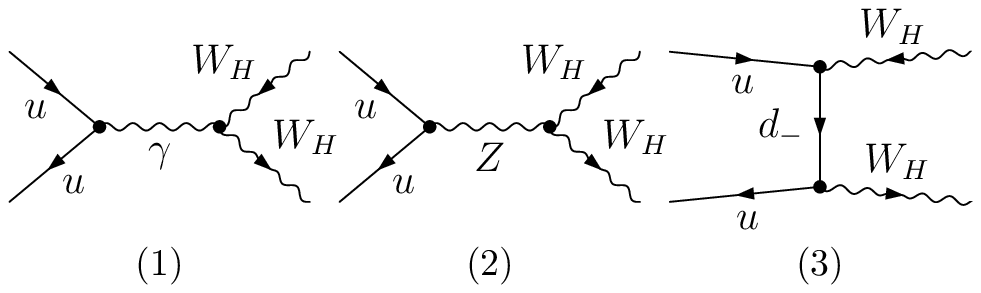}
\caption{ \label{fig1} The LO Feynman diagrams for the partonic
process \uuww. }
\end{center}
\end{figure}

The LO contribution to the cross section for the parent process
\ppww comes from the quark-antiquark annihilation. We denote the
subprocess as
\begin{eqnarray}
\label{process}
q(p_{1})+ \bar q(p_{2})\to W_H^+(p_{3})+W_H^-(p_{4}), &&
(q=u,d,c,s,b).
\end{eqnarray}
The corresponding Feynman diagrams for the $u\bar{u} \to W_H^+
W_H^-$ partonic process are shown in Fig.\ref{fig1}, the LO Feynman
diagrams for other partonic processes $q\bar{q} \to W_H^+
W_H^-$($q=d,c,s,b$) are similar with those in Fig.\ref{fig1} and are
not depicted there. Figs.\ref{fig1}(1,2) correspond to the exchanges
of $\gamma$ and $Z^0$ gauge bosons separately, and the diagram with
exchange of a $T$-odd quark is shown in Fig.\ref{fig1}(3). The
amplitudes of the tree-level Feynman diagrams in Figs.\ref{fig1} for
the partonic process $u\bar u \to W_H^+ W_H^-$ are respectively
expressed as
\begin{eqnarray}
{\cal M}_{u\bar u}^{(1)} &=& -\frac{2e^2}{3\hat s}\bar v
(p_2)\Big[(\rlap/ p_3-\rlap/ p_4)\epsilon^*(p_3)\cdot
\epsilon^*(p_4)-(2p_3+p_4)\cdot \epsilon^*(p_4) \rlap/
\epsilon^*(p_3)+(2p_4+p_3)
\cdot\epsilon^*(p_3)\rlap/ \epsilon^*(p_4)\Big]u(p_1), \nonumber\\
{\cal M}_{u\bar u}^{(2)} &=& -\frac{e^2}{2s^2_w (\hat s -m_Z^2)}
\bar v(p_2)  \Big[(\rlap/ p_3-\rlap/ p_4)\epsilon^*(p_3)\cdot
\epsilon^*(p_4)
-(2p_3+p_4)\cdot \epsilon^*(p_4) \rlap/ \epsilon^*(p_3) \Big. \nonumber\\
 &&+\Big. (2p_4+p_3)\cdot\epsilon^*(p_3)\rlap/ \epsilon^*(p_4)\Big] \cdot
 \left( P_L-\frac{4}{3}s^2_w \right) u(p_1), \nonumber \\
{\cal M}_{u\bar u}^{(3)} &=& -
\frac{e^2}{2s^2_w[(p_1-p_3)^2-m_{d_-}^2]}\bar v (p_2) \rlap/
\epsilon^*(p_4) P_L \Big[(\rlap/ p_1- \rlap/ p_3)+m_{d_-}\Big]\rlap/
\epsilon^*(p_3) P_L u(p_1), \label{M3}
\end{eqnarray}
where $P_L=\frac{1}{2}(1-\gamma_5)$. Analogously, we can get the
amplitudes for the other partonic processes $q\bar q \to W_H^+
W_H^-$ $(q=d,c,s,b)$. The LO amplitude for the partonic process
$q\bar q \to W_H^+ W_H^-$ can be generally expressed by summing up
all the above three total amplitudes,
\begin{eqnarray}
{\cal M}_{q\bar q}^{LO} = \sum_{i=1}^{3} {\cal M}_{q\bar q}^{(i)}, &&(q=u,d,c,s,b).
\end{eqnarray}
The LO cross section for the partonic process \qqww then can be
obtained as
\begin{eqnarray}
\hat{\sigma}^0_{q\bar q}= \frac{1}{4} \frac{1}{9}   \frac{(2 \pi
)^4}{4|\vec{p}|\sqrt{\hat{s}}}\int \sum_{spin}\sum_{color} |{\cal
M}^{LO}_{q\bar q}|^2 d\Omega_2,&&(q=u,d,c,s,b).
\end{eqnarray}
The factor $\frac{1}{4}$ and $\frac{1}{9}$ come from averaging over
the spins and colors of the initial partons respectively, $\vec{p}$
is the three-momentum of one initial parton in center-of-mass system
(CMS) and $\sqrt{\hat{s}}$ is the partonic CMS energy. The two-body
phase-space element $d\Omega_2$ is expressed as
\begin{eqnarray}
d\Omega_2 = \delta^{(4)}(p_1+p_2-p_3-p_4) \frac{d^3
\vec{p}_3}{(2\pi)^3 2E_3} \frac{d^3 \vec{p}_4}{(2\pi)^3 2E_4}.
\end{eqnarray}

\par
The total cross section for the parent process $pp \to W_H^+W_H^-+X$
at the tree-level can be obtained by integrating the cross section
for partonic processes $\hat{\sigma}^0_{q\bar q}$ with the parton
distribution functions (PDFs),
\begin{eqnarray} \label{pp-total cross section}\sigma_{LO}=\sum_{q=u,d}^{c,s,b} \int_0^1 dx_1
\int_0^1 dx_2 \left[ G_{q/P_1}(x_1,\mu_f)G_{\bar q/P_2}
(x_2,\mu_f)+(1 \leftrightarrow 2)
\right] \hat{\sigma}^0_{q\bar q}(\hat s = x_1 x_2 s),
\end{eqnarray}
where $G_{i/P}$ ($i=q,\bar q; P=P_1,P_2$) denotes the PDF of parton
$i$ in proton $P$, $x_{i}~(i=1,2)$ is the momentum fraction of a
parton in proton $P_{i}~(i=1,2)$, $\mu_f$ is the factorization scale
and $s$ is the total colliding energy squared in proton-proton CMS.

\par
\subsection{QCD NLO corrections  }
\par
The genuine QCD NLO correction to the parent process \ppww includes the
following components: (1) The QCD one-loop virtual corrections to
the partonic processes \qqww. (2) The contribution of the
real gluon emission partonic process \qqwwg . (3) The contribution
of the real light-(anti)quark emission partonic process \qgwwq. (4)
The corresponding collinear counterterms of the PDFs. In order to
isolate the ultraviolet (UV) and infrared (IR) singularities in the
NLO calculations, we adopt the dimensional regularization (DR)
method in $D=4-2\epsilon$ dimensions.

\par
\subsubsection{One-loop corrections to \qqww  partonic process }
\par
\begin{figure}
\begin{center}
\includegraphics[width=0.9\textwidth]{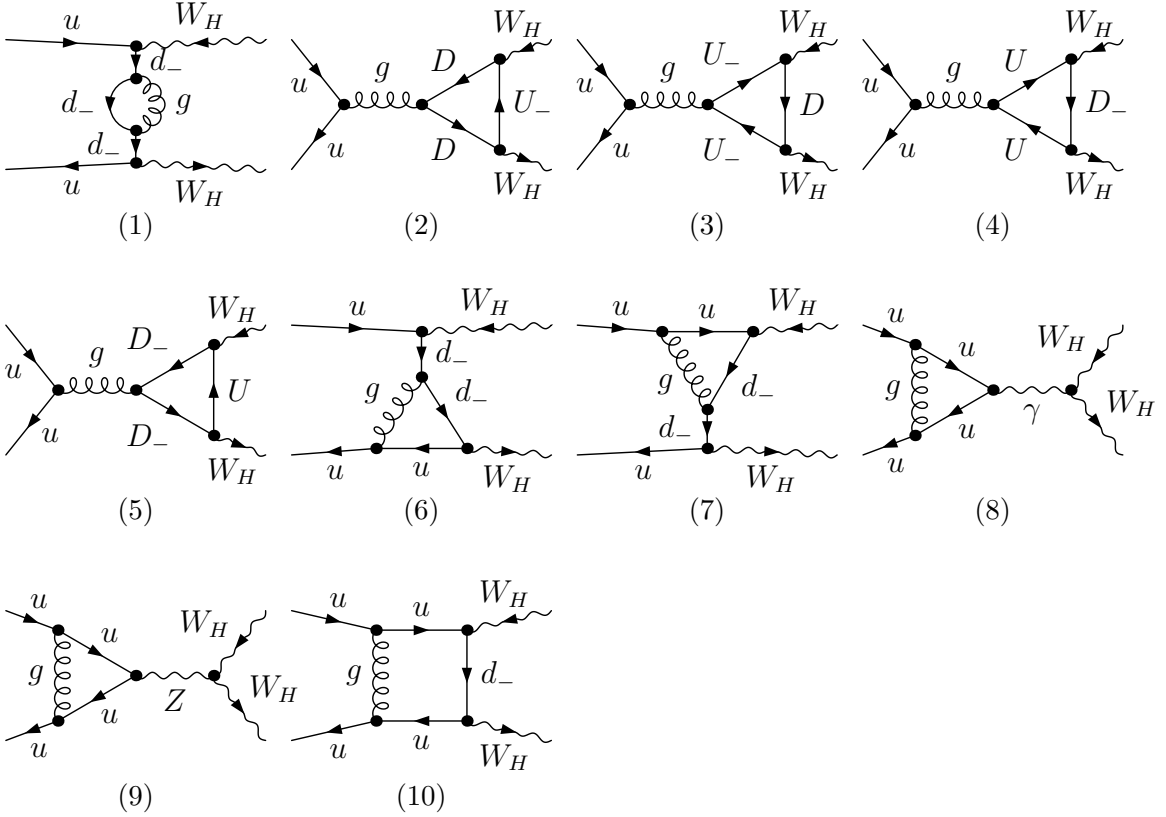}
\caption{ \label{fig2a} The representative one-loop Feynman diagrams
for the partonic process \uuww, where
$(U_-,D)=(u_-,d),(c_-,s),(t_-,b)$ and
$(U,D_-)=(u,d_-),(c,s_-),(t,b_-)$. }
\end{center}
\end{figure}

We plotted some representative Feynman diagrams for the one-loop
virtual corrections to the partonic process \uuww in
Fig.\ref{fig2a}. In calculating one-loop amplitudes we shall meet
both UV and IR singularities. In order to remove the UV divergences,
we renormalize the masses and wave functions of SM quarks and their
$T$-odd parteners with the counterterms defined as
\begin{eqnarray}
\psi^0_{q,L,R} & = & \left(1+\frac{1}{2}\delta Z_{q,L,R}\right)\psi_{q,L,R}~, \\
\psi^0_{q_-,L,R} & = & \left(1+\frac{1}{2}\delta Z_{q_-,L,R}\right)\psi_{q,L,R}~, \\
m^0_{q_-} & = & m_{q_-}+\delta m_{q_-}~,
\end{eqnarray}
where $\psi_{q,L,R},\psi_{q_-,L,R}$ denote the fields of SM quark
and $T$-odd quark respectively, and $m_{q_-}$denotes the mass of
$T$-odd quark. We adopt the on-shell scheme to perform the
renormalization procedure and then the relevant renormalization
constants are expressed as
\begin{eqnarray}
\delta Z_{q,L,R} & = & - \frac{\alpha_s (\mu_r)}{3\pi} \left[ \Delta_{UV} -
\Delta_{IR} \right]~, \\
\delta Z_{q_-,L,R} & = & - \frac{\alpha_s (\mu_r)}{3\pi} \left[
\Delta_{UV} + 2\Delta_{IR} + 4 + 3\ln
\left(\frac{\mu_r^2}{m_{q_-}^2}
\right) \right]~, \\
\frac{\delta m_{q_-}}{m_{q_-}} & = & - \frac{\alpha_s(\mu_r)}{3\pi}
\left\{3\left[\Delta_{UV}+\ln\left(\frac{\mu_r^2}{m_{q_-}^2}\right)\right]+4
\right\}~,
\end{eqnarray}
where $\Delta_{UV}=\frac{1}{\epsilon_{UV}}-\gamma_E + \ln (4\pi)$
and $\Delta_{IR}=\frac{1}{\epsilon_{IR}}-\gamma_E + \ln (4\pi)$.
After the renormalization, this one-loop virtual contribution
is UV finite. However, it still contains soft and
collinear IR singularities, which can be canceled by considering the
real gluon/light-(anti)quark emission subprocesses and the PDF
conterterms as described in the following subsections.

\par
\subsubsection{Real gluon/light-(anti)quark emission corrections}
\par
\begin{figure}
\begin{center}
\includegraphics[width=0.75\textwidth]{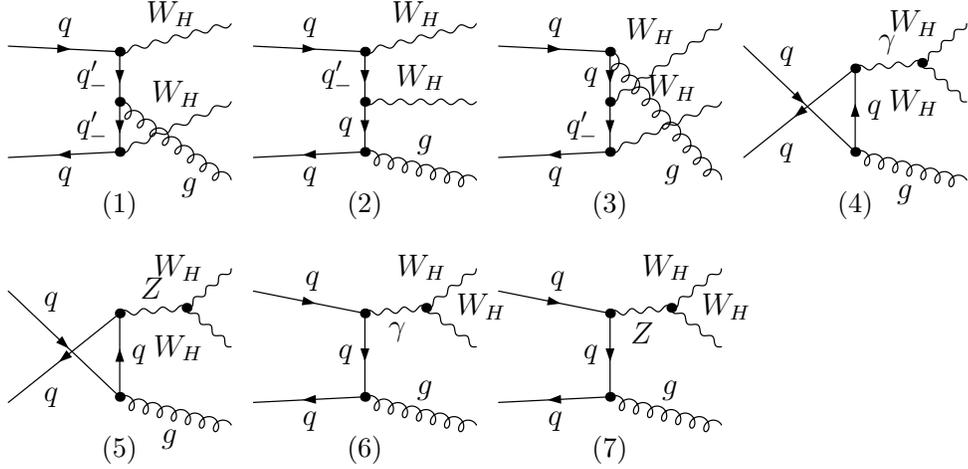}
\caption{ \label{fig2c} The tree-level Feynman diagrams for the real
gluon emission partonic process \qqwwg, $(q=u,d,c,s,b)$. }
\end{center}
\end{figure}

We denote the real gluon emission partonic process for the $W_H$-pair production as
\begin{eqnarray}
q(p_{1})+ \bar q(p_{2})\to W_H^+(p_{3})+W_H^-(p_{4})+g(p_{5}) , &&
(q=u,d,c,s,b).
\end{eqnarray}
The Feynman diagrams for this subprocess are shown in
Fig.\ref{fig2c}. There exist soft and collinear singularities in
these diagrams. In order to manipulate these IR divergences, we
employ the two cutoff phase-space slicing (TCPSS) methods \cite{19},
which introduce two arbitrary cutoff $\delta_s$ and $\delta_c$. The
soft cutoff $\delta_s$ divides the phase-space into two regions:
soft region ($E_5\leq \frac{1}{2}\delta_s \sqrt{\hat s}$) and hard
region ($E_5
> \frac{1}{2}\delta_s \sqrt{\hat s}$), another cutoff $\delta_c$
separates the hard region into hard collinear ($HC$) region ($\hat
s_{15}\leq \delta_c \hat s$ or $\hat s_{25}\leq \delta_c \hat s$)
and hard noncollinear ($\overline{HC}$) region.  Then we can express
the real gluon emission subprocess cross section as
\begin{eqnarray}
\hat{\sigma}_{g}= \hat{\sigma}_g^{S}+\hat{\sigma}_g^{HC}+
\hat{\sigma}_g^{\overline{HC}}.
\end{eqnarray}
The noncollinear cross section part $\hat{\sigma}_g^{\overline{HC}}$ is IR safe and the
soft singularity in the soft part $\hat{\sigma}_g^{S}$ can be canceled
by the soft IR divergence in the virtual corrections, as demonstrated by
the Kinoshita-Lee-Nauenberg (KLN) theorem \cite{KLN}. The collinear singularity can be
partially canceled by the virtual corrections, and the remained collinear divergence can be absorbed by the PDF counterterms.

\par
\begin{figure}
\begin{center}
\includegraphics[width=0.75\textwidth]{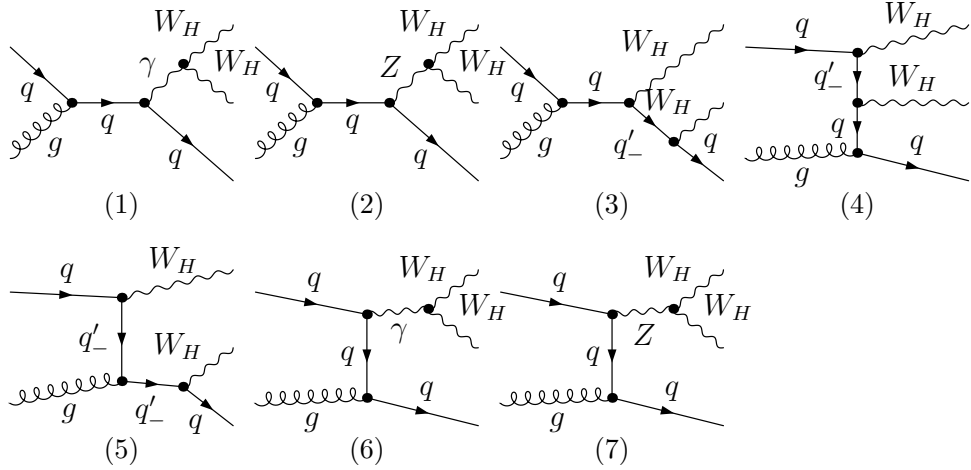}
\caption{ \label{fig3} The tree-level Feynman diagrams for real
light-quark emission partonic process $qg \to W_H^+W_H^-+q$,
$(q=u,d,c,s,b)$. }
\end{center}
\end{figure}

The real light-(anti)quark emission partonic process for the $W_H$-pair
production is denoted as
\begin{eqnarray}
q/\bar q(p_{1})+ g(p_{2})\to W_H^+(p_{3})+W_H^-(p_{4})+q/\bar q(p_{5}) , &&
(q=u,d,c,s,b).
\end{eqnarray}
We depict the tree-level Feynman diagrams for the real light-quark
emission partonic process $qg \to W_H^+W_H^-+q$ in Fig.\ref{fig3}.
We notice that there could exist resonance effect in Figs.\ref{fig3}(3)
and (5) due to possible one-shell $q_-$ propagator. In order to
deal with resonance singularity, we replace the $q_-$ mass squared $m_{q_-}^2$
in its propagator by $m_{q_-}^2-im_{q_-}\Gamma_{q_-}$. The expressions for the
partial decay widths of T-odd quarks are given
in Appendix B. This extremely large correction to Born
\ppww process would eventually destroy the perturbative convergence.
Furthermore, Figs.\ref{fig3}(3) and (5) are
included also in the $W_H q_-$ associated production process followed by
an on-shell decay $q_-\to W_H q'$. In order to avoid double counting
and to keep the convergence of the perturbative QCD description of
the $pp \to W_H^+W_H^- + X$ process, we need to remove the
intermediate on-shell $T$-odd quark $q_-$ contributions. This
removal can be implemented by adopting the PROSPINO subtraction
strategy \cite{PROSPINO-ref,on-shell subtraction}, which is done by
performing a replacement of the Breit-Wigner propagator:
\begin{eqnarray}
 \frac{|{\cal M}|^2( s_{W_H q} )}{( s_{W_H q} - m_{q_-}^2 )^2
 + m_{q_-}^2 \Gamma_{q_-}^2}
 & \to &
 \frac{|{\cal M}|^2( s_{W_H q} )}{( s_{W_H q} - m_{q_-}^2 )^2
 + m_{q_-}^2 \Gamma_{q_-}^2} \nonumber \\
 &&-
 \frac{|{\cal M}|^2( m_{q_-}^2 )}{( s_{W_H q} - m_{q_-}^2 )^2
 + m_{q_-}^2 \Gamma_{q_-}^2}
 \Theta( \hat{s} - 4 m_{q_-}^2 )
 \Theta( m_{q_-} - m_{W_H} ),
\end{eqnarray}
where $s_{W_H q}$ is the squared momentum flowing through the
intermediate $q_-$ propagator. For the real light-(anti)quark
emission corrections we use the cutoff $\delta_c$ to separate the
phase-space into collinear ($C$) region ($\hat s_{15}\leq \delta_c
\hat s$ or $\hat s_{25}\leq \delta_c \hat s$) and noncollinear
($\overline{C}$) region ($\hat s_{15} > \delta_c \hat s$ and $\hat
s_{25} > \delta_c \hat s$). Then we have
\begin{eqnarray}
\hat{\sigma}_{q(\bar{q})}= \hat{\sigma}_{q(\bar{q})}^{C}+
\hat{\sigma}_{q(\bar{q})}^{\overline{C}},
\end{eqnarray}
where $\hat{\sigma}_q^{\overline{C}}$ is finite and
$\hat{\sigma}_q^{C}$ contains collinear singularity. After summing
the virtual and real gluon/(anti)quark radiation corrections, the
remained collinear divergence can be canceled by that in the NLO
PDFs.

\par
\subsubsection{PDF counterterms}
\par
As mentioned above, part of the collinear divergences in the virtual
corrections to \qqww channel can be canceled by the real
gluon/light-(anti)quark emission partonic processes, and the
remained collinear divergences are absorbed in the PDF counterterms.
This collinear counterterm of the PDF can be denoted as $\delta
G_{i/P}(x,\mu_f)$ ($P=P_1,P_2$ and $i=g,u,\bar u,d,\bar d,c,\bar c,
s,\bar s, b,\bar{b}$ ). We divided $\delta G_{i/P}(x,\mu_f)$ into
two parts: the collinear gluon emission part $\delta
G_{i/P}^{(gluon)}(x,\mu_f)$ and the collinear light-quark emission
part $\delta G_{i/P}^{(quark)}(x,\mu_f)$,
\begin{eqnarray} \label{PDFcounterterm1}
&& \delta G_{q(g)/P}(x,\mu_f) = \delta G_{q(g)/P}^{(gluon)}(x,\mu_f)
                  +\delta G_{q(g)/P}^{(quark)}(x,\mu_f), \nb \\
&&~~~~~~~~~~~~~~~~(q = u, \bar{u}, d, \bar{d}, c, \bar{c}, s,
\bar{s}, b,\bar{b} ).
\end{eqnarray}
These PDF counterterms can be expressed as
\begin{eqnarray} \label{PDFcounterterm2}
&& \delta G_{q(g)/P}^{(gluon)}(x,\mu_f) =
   \frac{1}{\epsilon} \left[
                      \frac{\alpha_s}{2 \pi}
                      \frac{\Gamma(1 - \epsilon)}{\Gamma(1 - 2 \epsilon)}
                      \left( \frac{4 \pi \mu_r^2}{\mu_f^2} \right)^{\epsilon}
                      \right]
   \int_x^1 \frac{dz}{z} P_{qq(gg)}(z) G_{q(g)/P}(x/z,\mu_f), \nonumber \\
&& \delta G_{q/P}^{(quark)}(x,\mu_f) =
   \frac{1}{\epsilon} \left[
                      \frac{\alpha_s}{2 \pi}
                      \frac{\Gamma(1 - \epsilon)}{\Gamma(1 - 2 \epsilon)}
                      \left( \frac{4 \pi \mu_r^2}{\mu_f^2} \right)^{\epsilon}
                      \right]
   \int_x^1 \frac{dz}{z} P_{qg}(z) G_{g/P}(x/z,\mu_f),  \nonumber \\
&& \delta G_{g/P}^{(quark)}(x,\mu_f) =
   \frac{1}{\epsilon} \left[
                      \frac{\alpha_s}{2 \pi}
                      \frac{\Gamma(1 - \epsilon)}{\Gamma(1 - 2 \epsilon)}
                      \left( \frac{4 \pi \mu_r^2}{\mu_f^2} \right)^{\epsilon}
                      \right]
   \sum_{q=u,\bar{u},d,\bar{d},}^{c, \bar {c}, s, \bar {s},b,\bar{b},}
   \int_x^1 \frac{dz}{z} P_{gq}(z) G_{q/P}(x/z,\mu_f),~~~
\end{eqnarray}
where $P_{ij}(z) (ij=qq,qg,gq,gg)$ denote the splitting functions.
One can find their explicit expressions in Ref.\cite{19}.

\par
\subsubsection{Correction from \ggww partonic process }
\par
\begin{figure}
\begin{center}
\includegraphics[width=0.7\textwidth]{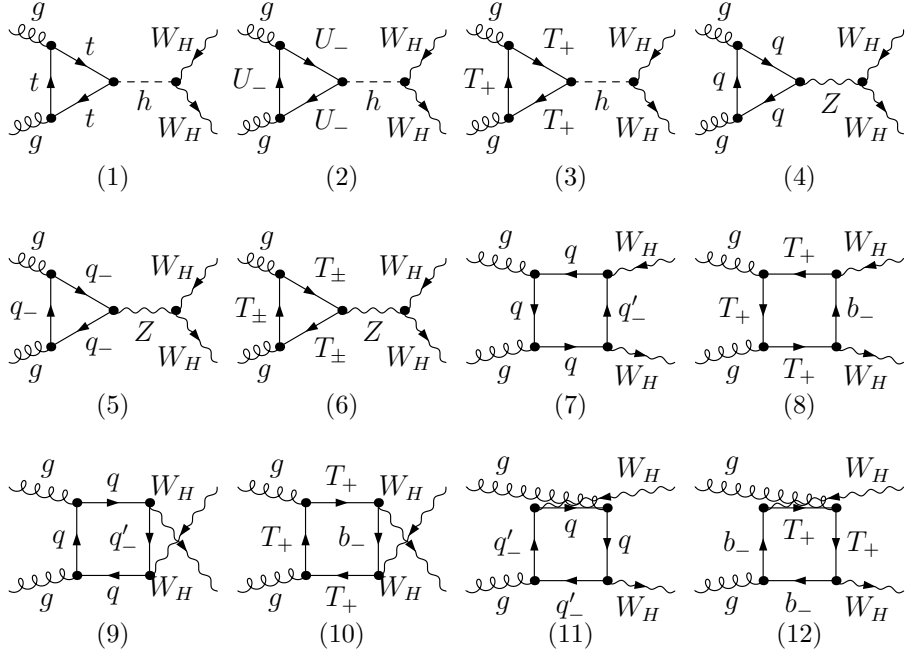}
\caption{ \label{fig2b} The representative lowest order Feynman
diagrams for the partonic process \ggww, where $U_-=u_-,c_-,t_-$,
$q,q'=u,d,c,s,b,t$ and $q_-,q'_-=u_-,d_-,c_-,s_-,b_-,t_-$. }
\end{center}
\end{figure}

The gluon-gluon fusion partonic process \ggww contributes also to the
parent process $pp\to W_H^+W_H^-+X$.
We can see that the QCD NLO correction to the partonic process \qqww is at the order of
$\alpha_{ew}^2\alpha_{s}$, while the lowest order partonic process
\ggww is at the order of $\alpha_{ew}^2\alpha_{s}^2$. The later LO contribution is
$\alpha_{s}$ order higher than the QCD NLO contribution from  previous
subprocess. But both contribution parts might be comparable with each other due to the
large gluon luminosity at the TeV-scale collider LHC.
The representative Feynman diagrams for gluon-gluon fusion partonic
process are depicted in Fig.\ref{fig2b}. The total one-loop
amplitude ${\cal M}^{1-loop}_{gg}$ for this partonic process is UV
and IR finite, and the cross section at the lowest order,
$\hat{\sigma}^0_{gg}$, can be expressed as
\begin{eqnarray}
\hat{\sigma}^0_{gg}= \frac{1}{4} \frac{1}{64}   \frac{(2 \pi
)^4}{4|\vec{p}|\sqrt{\hat{s}}}\int \sum_{spin}\sum_{color} |{\cal
M}^{1-loop}_{gg}|^2 d\Omega_2.
\end{eqnarray}

\par
The total cross section for the parent process $pp \to gg \to W_H^+W_H^-+X$
at the lowest order can be obtained by integrating the cross section
for partonic process $\hat{\sigma}^0_{gg}$ with the gluon PDF in proton $G_{g/P}(x, \mu)$,
\begin{eqnarray} \label{pp-gg-total cross section}
\sigma(pp \to gg \to W_H^+W_H^-+X)= \frac{1}{2}\int_0^1 dx_1
\int_0^1 dx_2 \left[ G_{g/P_1}(x_1,\mu_f)G_{g/P_2}
(x_2,\mu_f)+(1 \leftrightarrow 2)
\right] \hat{\sigma}^0_{gg}(\hat s = x_1 x_2 s), && \nb \\
\end{eqnarray}
where we adopt the notations same as in Eq.(\ref{pp-total cross section}).

\par
\subsubsection{Total QCD NLO correction }
\par
After the renormalization and summing up all the QCD NLO one-loop
corrections, the gluon/light-(anti)quark emission corrections and
the PDF counterterm contributions, all the UV and IR (both soft and
collinear) singularities are eliminated. Consequently, the QCD NLO
corrected integrated cross section for the \ppww process is finite
and can be expressed as
\begin{eqnarray}\label{TotalSigmaCorr}
\sigma_{NLO}&=&\sigma_{LO}+\Delta
\sigma_{NLO}= \sigma_{LO}+\Delta\sigma^{(2)}+\Delta\sigma^{(3)} \nb \\
&=&\sigma_{LO}+\Delta\sigma_{NLO}(pp \to q\bar{q} \to W_H^+W_H^-+X)
+\sigma(pp \to gg \to W_H^+W_H^-+X).
\end{eqnarray}
where $\Delta\sigma_{NLO}$ consists of two parts, $\Delta\sigma_{NLO}(pp \to q\bar{q}
\to W_H^+W_H^-+X)$ and $\sigma(pp \to gg \to W_H^+W_H^-+X)$.
The two-body term $\Delta \sigma^{(2)}$ includes the
one-loop corrections to the $pp \to q\bar{q} \to W_H^+W_H^-+X$ process,
the lowest order contribution from the $pp \to gg \to W_H^+W_H^-+X$
process and the cross sections for the real
gluon/light-(anti)quark emission processes over the soft and hard
collinear phase-space regions, while the three-body term $\Delta
\sigma^{(3)}$ contains the cross sections for the real
gluon/light-(anti)quark emission processes over the hard
noncollinear regions.

\vskip 5mm
\section{Numerical results and discussions }\label{numres}
\par
\subsection{Input parameters}\label{parameters}
\par
As discussed in Ref.\cite{Blanke:2007ckm}, the two mixing matrices
satisfy $V_{Hu}^{\dag}V_{Hd}=V_{CKM}$. Therefore, they cannot
simultaneously be set to the identity. In the following calculations
we take $V_{Hu}$ to be a unit matrix, then we have $V_{Hd}=V_{CKM}$.
We take $\alpha_{ew}(m_Z^2)^{-1}=127.916$, $m_W=80.399~GeV$,
$m_Z=91.1876~GeV$,
$\sin^2\theta_W=1-\left(\frac{m_W}{m_Z}\right)^2=0.2226$ and
$m_t=171.2~GeV$ \cite{databook}. The masses of all the SM leptons
and quarks except top quark are neglected. The center-of-mass
energies $\sqrt{s}$ of proton-proton collision are taken to be
$14~TeV$ and $8~TeV$ for the future and early LHC, separately. We
set the factorization and renormalization scale to be equal ($\mu_r
= \mu_f$) and define $\mu_0=m_{W_H}$. We employ CTEQ6L1 and CTEQ6M
in the the LO and NLO calculations respectively \cite{cteq}, and fix the LHT
parameters $\kappa=1$ and $s_\alpha = c_\alpha =
\frac{\sqrt{2}}{2}$. Then the masses of heavy gauge bosons and
$T$-odd quarks are only the functions of the LHT parameter $f$ as
shown in Eqs.(\ref{m_v}) and (\ref{m_q}). The
Cabibbo-Kobayashi-Maskawa (CKM) matrix elements are taken as
\begin{eqnarray}\label{CKM}
 V_{CKM} &=& \left(
\begin{array}{ccc}
    V_{ud} \ &  V_{us} \ &  V_{ub} \\
    V_{cd} \ &  V_{cs} \ &  V_{cb} \\
    V_{td} \ &  V_{ts} \ &  V_{tb} \\
\end{array}
    \right)=\left(
\begin{array}{ccc}
     0.97418 \ &  0.22577 \ &  0 \\
    -0.22577 \ &  0.97418 \ &  0 \\
       0 \ &  0 \ &  1 \\
\end{array}  \right).
\end{eqnarray}

\par
\subsection{Masses and Decay widths }
\par
From Eq.(\ref{m_q}) we can see that all the $T$-odd quarks
$q_-$ $(q_- = u_-, d_-, s_-, c_-, b_-, t_-)$ have nearly equal
masses when the scale
$f$ is large enough. By using Eqs.(\ref{m_v}), (\ref{m_q}),
(\ref{T-oddMass}) and (\ref{t&T-evenMass}), and taking the LHT
parameters $\kappa = 1$ and $s_\alpha = c_\alpha =
\frac{\sqrt{2}}{2}$, we obtain the masses of heavy gauge
bosons, $T$-odd quarks and the quarks in top sector for some typical values of the LHT
global symmetry breaking scale $f$, and list them in Table \ref{tab1}.
\begin{table}
\begin{center}
\begin{tabular}{c|c|c|c|c|c}
  \hline
    $f$    &  $m_{W_H}\approx m_{Z_H}$ & $m_{u_-}=m_{c_-}=m_{t_-}$  & $m_{d_-}=
m_{s_-}=m_{b_-}$ & $m_{T_+}$ & $m_{T_-}$ \\
   $~~(GeV)~~$ & $~~(GeV)~~$  & $~~(GeV)~~$ & $~~(GeV)~~$ & $~~(GeV)~~$ & $~~(GeV)~~$ \\
  \hline
     500  & 322.1  & 685.7   & 707.1   & 695.9  & 507.0   \\
     700  & 457.8  & 974.7   & 989.9   & 974.3  & 699.6   \\
     800  & 525.1  & 1118.0  & 1131.4  & 1113.5 & 796.7   \\
     900  & 592.3  & 1260.9  & 1272.8  & 1252.7 & 894.1   \\
     1100 & 726.1  & 1545.9  & 1555.6  & 1531.1 & 1089.4  \\
     1300 & 859.7  & 1830.3  & 1838.5  & 1809.4 & 1285.2  \\
     1500 & 993.1  & 2114.2  & 2121.3  & 2087.8 & 1481.3  \\
  \hline
\end{tabular}
\end{center}
\begin{center}
\begin{minipage}{15cm}
\caption{\label{tab1} The masses of $W_H$, $q_-$
($q_-=u_-,d_-,c_-,s_-,b_-,t_-$) and $T_{\pm}$ for some typical
values of the LHT parameter $f$ with $\kappa =1$ and $s_\alpha =
c_\alpha = \frac{\sqrt{2}}{2}$. }
\end{minipage}
\end{center}
\end{table}

\par
\begin{figure}
\begin{center}
\includegraphics[width=0.65\textwidth]{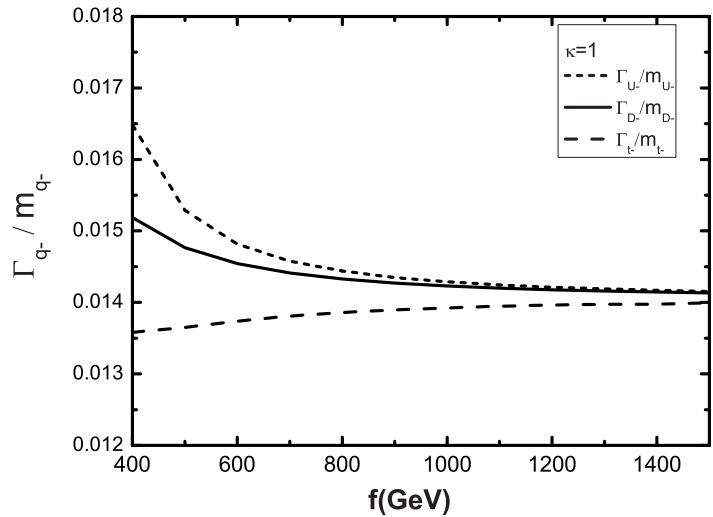}
\caption{ \label{fig4} The LO total decay widths of the $T$-odd quarks as
functions of the global symmetry breaking scale $f$, where
$q_-=U_-,D_-,t_-$ and $(U_-,D_-)=(u_-,d_-),(c_-,s_-)$. }
\end{center}
\end{figure}

With above related parameters and Eqs.(\ref{Width-1})
and (\ref{Width-2}) in Appendix B, we obtain numerically the LO decay widths
$\Gamma_{q_-}$ as functions of $f$ in Fig.\ref{fig4},
where we give only the curves for $U_-=u_-,c_-$, $D_-=d_-,s_-$ and
$t_-$-quarks. The decay width of the $T$-odd $b_-$-quark is not
presented in the figure, since it is not relevant in our calculations.

\par
\subsection{Checks}
\par
The correctness of our calculations are verified in the following aspects:
\par
{\bf 1.} We adopt the same input parameters and PDFs as used in
Ref.\cite{cpyuan:2006ph}, and compare our LO results with those as
shown in Fig.9 of Ref.\cite{cpyuan:2006ph}. We find that our LO
cross sections are in good agreement with those read out from the
figure.

\par
{\bf 2.} The cancelations of UV and IR divergences are verified numerically after
combining all the contributions at the QCD NLO.

\par
{\bf 3.} The one-loop virtual corrections are computed independently
by using two different programs. One is based on our in-house codes for
numerical evaluation of the one-loop integrals, and the other
is LoopTools 2.2 package. We find that the numerical results coincide
with each other within the calculation errors.

\par
\begin{figure}[htbp]
\begin{center}
\includegraphics[scale=0.7]{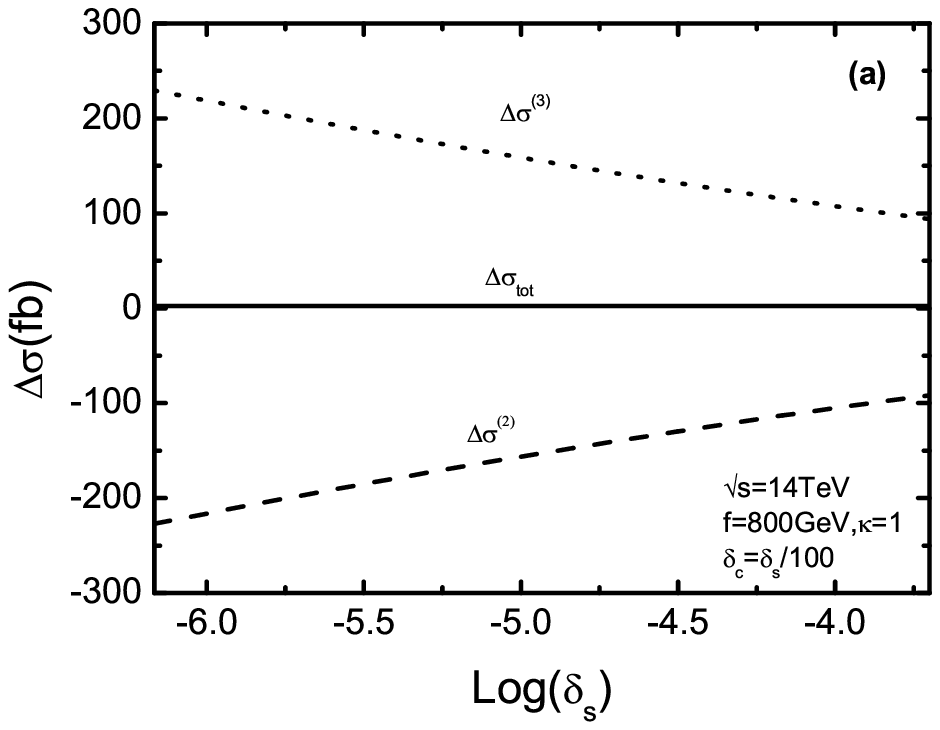}
\includegraphics[scale=0.7]{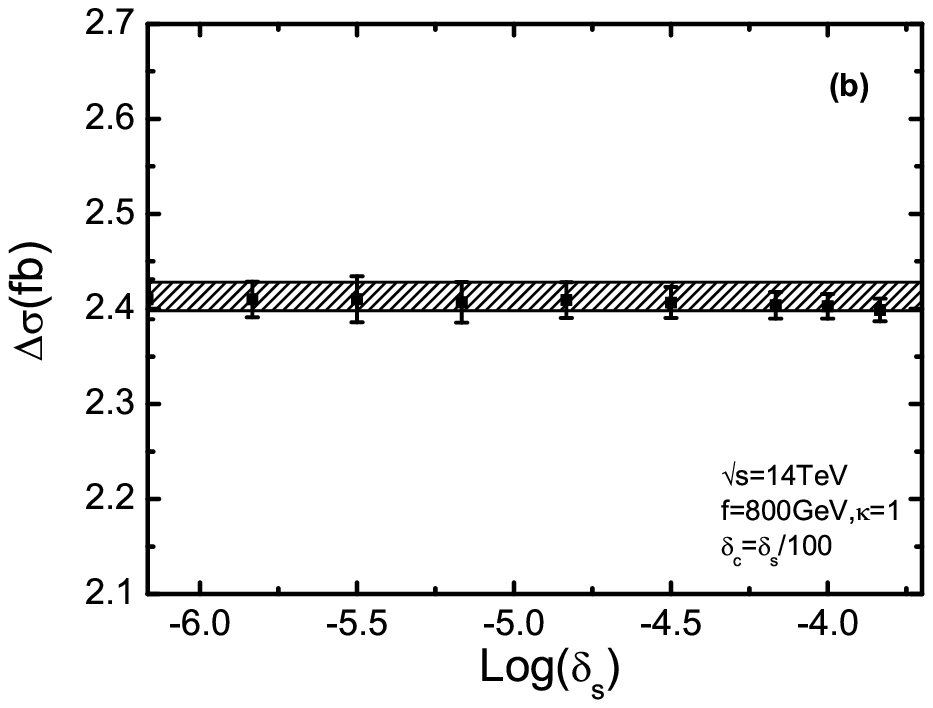}
\hspace{0in}%
\caption{\label{fig5} (a) The dependence of the QCD NLO corrections
to the $pp \to u\bar u \to W_H^+ W_H^- + X$ process on the cutoffs
$\delta_s$ and $\delta_c$ at the $\sqrt{s}=14~TeV$ LHC, where we take
$f=800~GeV$, $\kappa=1$, $s_\alpha = c_\alpha = \frac{\sqrt{2}}{2}$,
$\delta_c=\delta_s/100$ and $\mu=\mu_0=m_{W_H}=525.15~GeV$.
(b) The amplified curve for $\Delta\sigma_{tot}$ in Fig.\ref{fig5}(a).
The shadowing region shows the result by adopting DPS method with
$\pm 1\sigma$ statistic error. }
\end{center}
\end{figure}

{\bf 4.} The $\delta_s/\delta_c$ independence of the total QCD NLO
correction has been numerically verified. As mentioned above, in the
TCPSS method two arbitrary cutoffs are introduced to separate the
phase-space in order to isolate the IR divergences. From
Eq.(\ref{TotalSigmaCorr}), the total QCD NLO correction ($\Delta
\sigma_{NLO}$) can be divided into the two-body and three-body
corrections ($\Delta \sigma^{(2)}$ and $\Delta \sigma^{(3)}$) by the
two cutoffs. We depict $\Delta \sigma^{(2)}$, $\Delta \sigma^{(3)}$
and $\Delta \sigma_{NLO}$ for the process $pp \to u \bar{u} \to
W_H^+W_H^-+X$ as functions of the soft cutoff $\delta_s$ in
Fig.\ref{fig5}(a) with $f=800~GeV$, $\kappa=1$, $s_\alpha = c_\alpha
= \frac{\sqrt{2}}{2}$, $\delta_c=\delta_s/100$ and
$\mu=\mu_0=m_{W_H}=525.15~GeV$. The amplified curve for the total
correction $\Delta\sigma_{NLO}$ in Fig.\ref{fig5}(a) is demonstrated
in Fig.\ref{fig5}(b) together with calculation errors. We adopt also
the dipole subtraction (DPS) method \cite{dipole} to deal with the
IR singularities for further verification. The $\Delta \sigma_{NLO}$
result by adopting this method with $\pm 1\sigma$ statistic error is
plotted as the shadowing region in Fig.\ref{fig5}(b). We can see
that the results from both the TCPSS method and the DPS method are
in good agreement. From these two figures we find that the total QCD NLO
correction $\Delta \sigma_{NLO}$ is independent of the two
cutoffs within the statistical errors. This independence is an
indirect check for the correctness of our work. In further numerical
calculations, we fix $\delta_s = 1\times 10^{-4}$ and
$\delta_c=1\times 10^{-6}$.

\par
\subsection{Dependence on factorization/renormalization scale  }
\par
\begin{figure}[htbp]
\begin{center}
\includegraphics[width=0.45\textwidth]{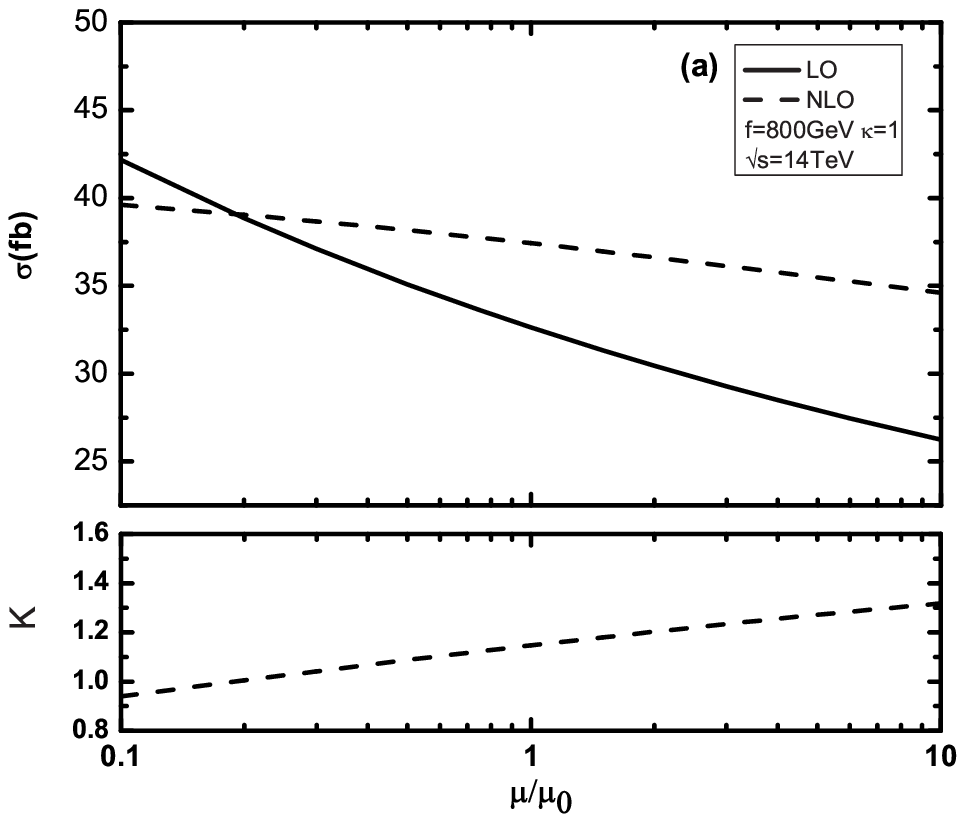}
\includegraphics[width=0.45\textwidth]{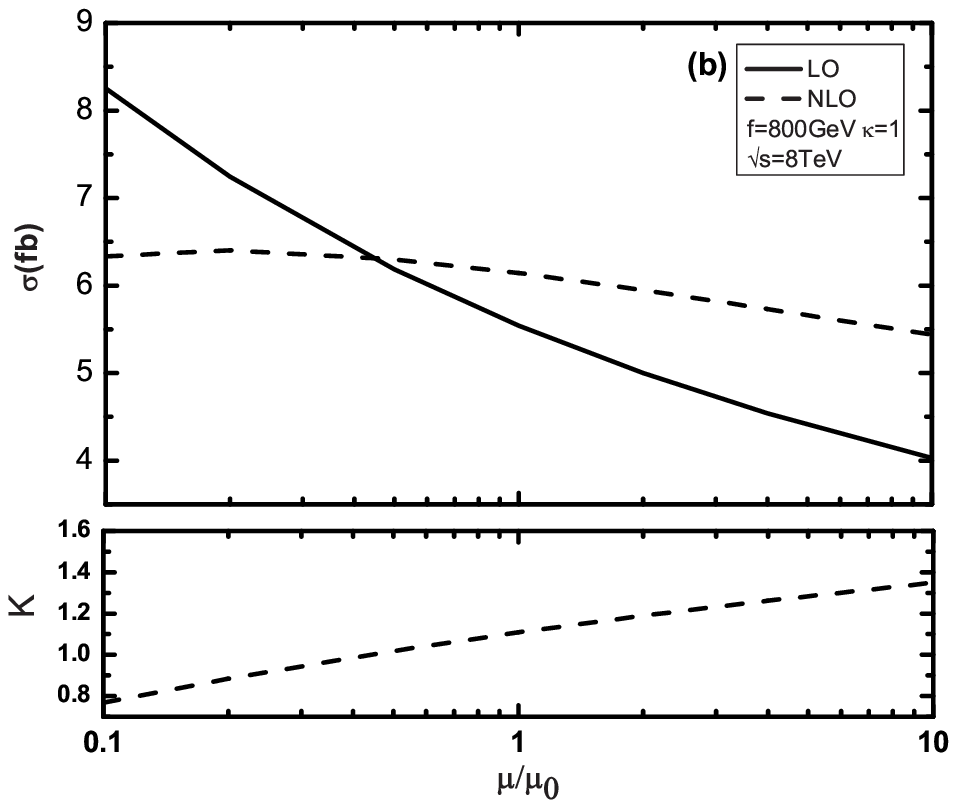}
\hspace{0in}%
\caption{\label{fig6} The dependence of the LO, QCD NLO corrected integrated
cross sections and the corresponding $K$-factors for the \ppww process on the
factorization/renormalization scale $\mu$ at the LHC with $f=800~GeV$,
$\kappa=1$ and $s_\alpha = c_\alpha = \frac{\sqrt{2}}{2}$.
(a) $\sqrt{s}=14~TeV$. (b) $\sqrt{s}=8~TeV$. }
\end{center}
\end{figure}

In Figs.\ref{fig6}(a,b) we present the dependence of the LO, QCD NLO
corrected integrated cross sections and the corresponding $K$-factor
($K\equiv \sigma_{NLO}/\sigma_{LO}$) on the
factorization/renormalization scale $\mu$ for the process \ppww at
the $\sqrt{s}=14~TeV$ and the $\sqrt{s}=8~TeV$ LHC separately, where
we take the LHT parameters $f=800~GeV$, $\kappa = 1$ and $s_\alpha =
c_\alpha = \frac{\sqrt{2}}{2}$. From the curves in
Figs.\ref{fig6}(a,b), we find that QCD NLO corrections to the $pp
\to W_H^+ W_H^- + X$ process significantly reduce the scale
uncertainty. We can read out from the figures that the LO and QCD
NLO corrected cross sections at $\mu_0=m_{W_H}$ are
$\sigma_{LO}(\sqrt{s}=14TeV)=32.63^{+9.56}_{-6.38} ~fb$,
$\sigma_{NLO}(\sqrt{s}=14TeV)=37.43^{+2.19}_{-2.83} ~fb$ and
$\sigma_{LO}(\sqrt{s}=8TeV)=5.54^{+2.71}_{-1.51} ~fb$,
$\sigma_{NLO}(\sqrt{s}=8TeV)=6.14^{+0.26}_{-0.70} ~fb$, where the
uncertainties describe the missing higher-order corrections
estimated via scale variations in the range of $0.1 \mu_0 < \mu <
10\mu_0$. The $K$-factor varies from $0.94~(0.77)$ to $1.32~(1.35)$
at the $\sqrt{s}=14~TeV$ ($8~TeV$) LHC, when $\mu/\mu_0$ goes from
$0.1$ to $10$. With the definition of scale uncertainty as
$\eta=\frac{|\sigma(0.1\mu_0)-\sigma(10\mu_0)|}{\sigma(\mu_0)}$, we
obtain that the scale uncertainties are reduced from $48.88\%$ (LO)
to $13.40\%$ (NLO) at the $\sqrt{s}=14~TeV$ and from $76.23\%$ (LO)
to $14.54\%$ (NLO) at the $\sqrt{s}=8~TeV$ LHC, respectively. In
Table \ref{tab1-1} we list some numerical results of the cross
sections and $K$-factors for some typical values of $\mu/\mu_0$,
which are read out from Figs.\ref{fig6}(a,b). In order to
investigate the contribution from the $pp \to gg \to W_H^+W_H^-+X$
process, which is considered as a component of the QCD NLO
corrections to the parent process \ppww, we also present the cross
sections for the $pp \to gg \to W_H^+W_H^-+X$ process ($\sigma(gg)$)
in this table. We can obtain from the data that the QCD NLO
correction part from the $pp \to gg \to W_H^+W_H^-+X$ process at
$\mu=\mu_0$ is about $5.85\%$ ($3.23\%$) of the total QCD NLO
correction ($\Delta \sigma_{NLO}$) at the $14~TeV~(8~TeV)$ LHC. We
can see also that the NLO theoretical uncertainty due the choice of
$\mu$ mainly comes from the genuine QCD NLO corrected cross section
for the $pp \to q\bar{q} \to W_H^+W_H^-+X$ process, while the
contribution from the $pp \to gg \to W_H^+W_H^-+X$ process is
relatively small. In further numerical calculations we fix the
renormalization and factorization scales being equal to their
central value, i.e., $\mu=\mu_r=\mu_f=\mu_0=m_{W_H}$.
\begin{table}
\begin{center}
\begin{tabular}{c|c|c|c|c|c}
\hline
$\sqrt{s}$&$\mu/\mu_0$&$\sigma_{LO}$&$\sigma_{NLO}$&$\sigma(gg)$&$K$\\
$(TeV)$& &$(fb)$&$(fb)$&$(fb)$&\\
\hline
 &0.1&42.190(1)&39.62(2)&0.993(1)&0.939\\
 &0.5&35.091(1)&38.17(2)&0.3946(6)&1.09\\
14&1&32.626(1)&37.43(2)&0.2810(5)&1.15\\
 &2&30.444(1)&37.53(2)&0.2056(4)&1.20\\
 &10&26.242(1)&34.60(2)&0.1081(2)&1.32\\
\hline
 &0.1&8.2548(4)&6.333(3)&0.0805(1)&0.767\\
 &0.5&6.1850(3)&6.300(3)&0.02842(3)&1.019\\
8&1&5.5417(2)&6.143(3)&0.01943(3)&1.11\\
 &2&5.0006(2)&5.947(3)&0.01371(2)&1.21\\
 &10&4.0304(2)&5.440(2)&0.00671(1)&1.40\\
\hline
\end{tabular}
\end{center}
\begin{center}
\begin{minipage}{15cm}
\caption{\label{tab1-1} The numerical results of $\sigma_{LO}$,
$\sigma_{NLO}$ and the corresponding $K$-factors at the
$14~TeV$ and the $8~TeV$ LHC by taking $f=800~GeV$, $\kappa =
1$, $s_\alpha = c_\alpha = \frac{\sqrt{2}}{2}$ and some typical
values of factorization/renormalization scale $\mu$.
$\sigma(gg)$ is the cross section for the
$pp \to gg \to W_H^+W_H^-+X$ process, which is considered as a component
of the QCD NLO correction to the parent process \ppww. }
\end{minipage}
\end{center}
\end{table}

\par
\subsection{Dependence on global symmetry breaking scale $f$ }
\par
\par
\begin{figure}[htbp]
\begin{center}
\includegraphics[width=0.45\textwidth]{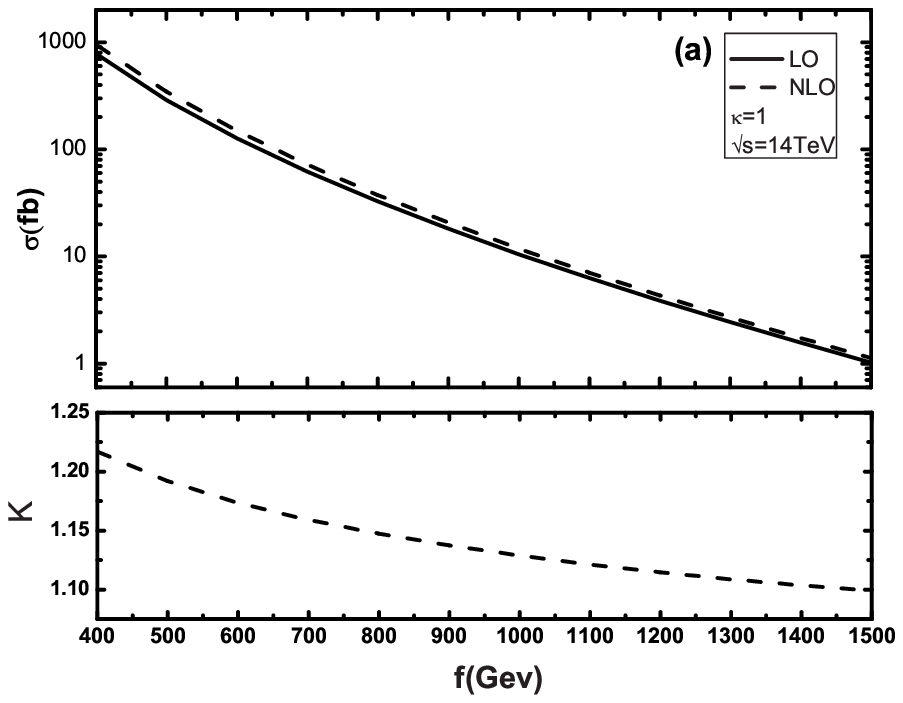}
\includegraphics[width=0.45\textwidth]{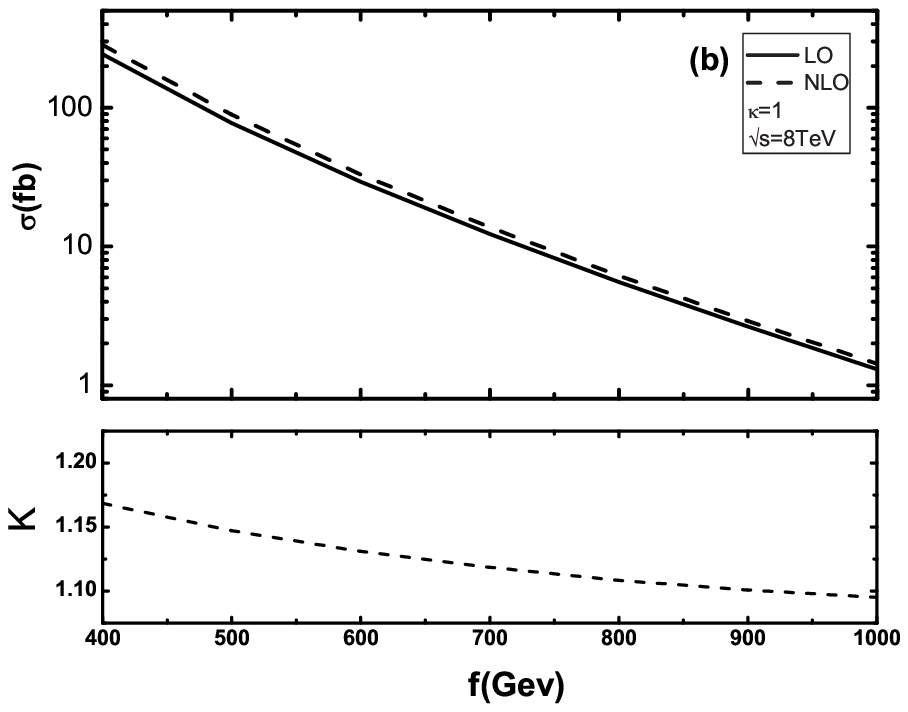}
\hspace{0in}%
\caption{\label{fig7} The LO, QCD NLO corrected integrated cross sections
and the corresponding $K$-factors for the \ppww process as the functions
of the global symmetry breaking scale $f$ at the LHC with $\kappa =1$ and
$s_\alpha = c_\alpha = \frac{\sqrt{2}}{2}$. (a) $\sqrt{s}=14~TeV$.
(b) $\sqrt{s}=8~TeV$. }
\end{center}
\end{figure}

The LO and QCD NLO corrected integrated cross sections together with
the corresponding $K$-factor as functions of the scale $f$ at the
$\sqrt{s}=14~TeV$ and the $\sqrt{s}=8~TeV$ LHC are depicted in
Figs.\ref{fig7}(a) and (b), respectively. We can see from
Fig.\ref{fig7} that the LO and NLO total cross sections for the
\ppww process decrease drastically when $f$ goes up. This is because
the mass of final $W_H$ becomes heavier as the increment of $f$,
therefore the phase-space becomes smaller. We can read out from the
figures that the corresponding $K$-factor varies from $1.22$ to
$1.10$ at the $\sqrt{s}=14~TeV$ LHC and from $1.17$ to $1.10$ at the
$\sqrt{s}=8~TeV$ LHC in the plotted $f$ range. In Table \ref{tab3}, we
list some numerical results of the LO, NLO cross sections and the
corresponding $K$-factors for some typical values of $f$ which are
shown in Figs.\ref{fig7}(a,b).
\begin{table}
\begin{center}
\begin{tabular}{c|c|c|c|c}
\hline
$\sqrt{s}$  & $f$ & $\sigma_{LO}$  & $\sigma_{NLO}$ & $K$  \\
$(TeV)$ & $(GeV)$ & $(fb)$ & $(fb)$ &  \\
\hline
     & 500  & 289.93(1)    &345.6(2)      &1.19    \\
     & 700  & 62.053(3)    &71.93(4)      &1.16    \\
     & 800  & 32.626(1)    &37.43(2)      &1.15    \\
 14  & 900  & 18.1252(8)   &20.61(1)      &1.14    \\
     & 1100 & 6.2964(3)    &7.059(3)      &1.12    \\
     & 1300 & 2.44312(9)   &2.709(1)      &1.11    \\
     & 1500 & 1.02314(4)   &1.1246(5)     &1.10    \\
\hline
    & 500  & 77.693(3)     &89.12(5)      &1.15    \\
  8 & 700  & 12.2863(5)    &13.741(7)     &1.12    \\
    & 800  & 5.5417(2)     &6.143(3)      &1.11    \\
    & 900  & 2.6357(1)     &2.901(1)      &1.10    \\
\hline
\end{tabular}
\end{center}
\begin{center}
\begin{minipage}{15cm}
\caption{\label{tab3} The numerical results of $\sigma_{LO}$,
$\sigma_{NLO}$ and the corresponding $K$-factors at the
$\sqrt{s}=14~TeV$ and the $\sqrt{s}=8~TeV$ LHC by taking $\kappa =
1$, $s_\alpha = c_\alpha = \frac{\sqrt{2}}{2}$, $\mu=\mu_0$ and some typical
values of $f$.  }
\end{minipage}
\end{center}
\end{table}

\par
\subsection{Differential cross sections }
\par
\begin{figure}[htbp]
\begin{center}
\includegraphics[width=0.45\textwidth]{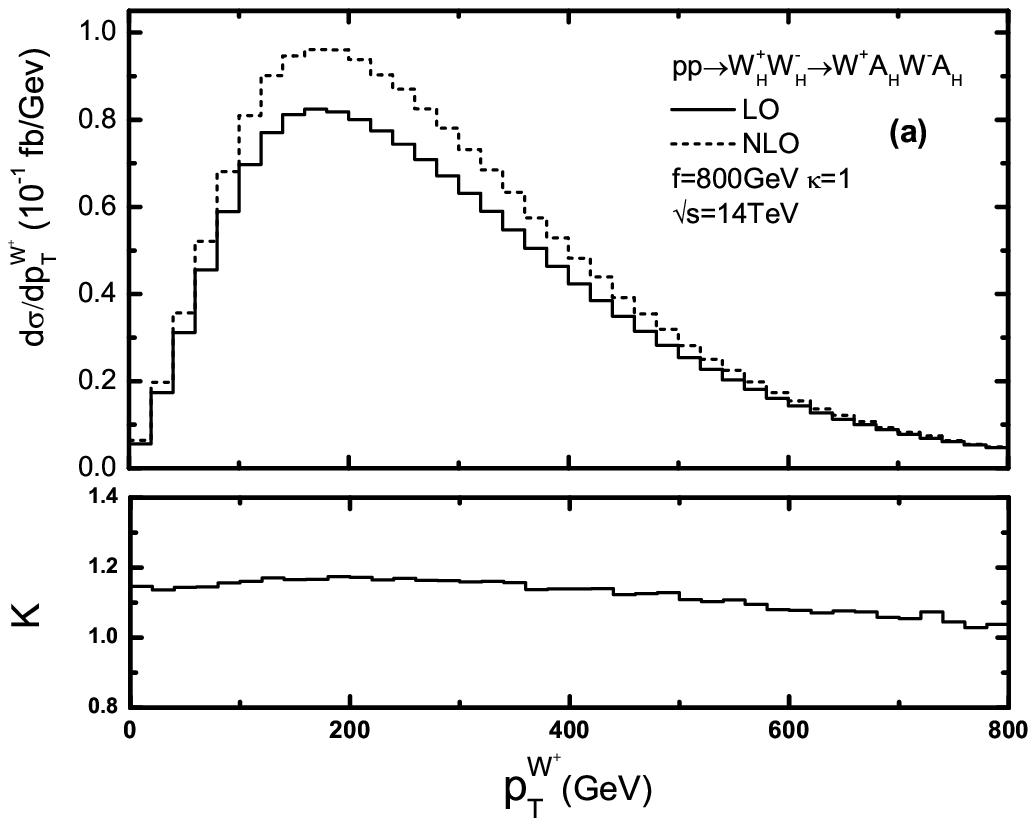}
\includegraphics[width=0.45\textwidth]{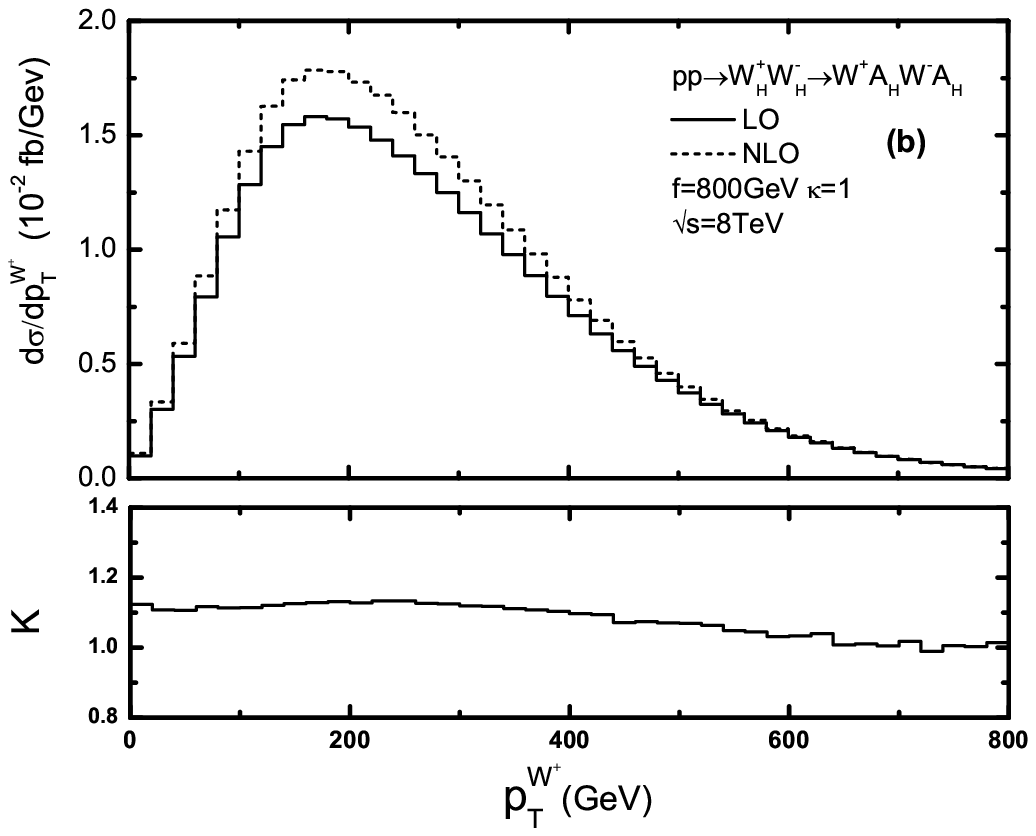}
\caption{\label{fig8} The LO, QCD NLO corrected $p_T^{W^+}$
distributions and the corresponding $K$-factors of final $W^+$
boson for the \ppww process at the LHC by taking $f=800~GeV$,
$\kappa=1$ and $s_\alpha = c_\alpha = \frac{\sqrt{2}}{2}$. (a) $\sqrt{s}=14~TeV$.
(b) $\sqrt{s}=8~TeV$. }
\end{center}
\end{figure}

In this subsection, we investigate the kinematic distributions of
final products after the subsequential decays of heavy charged gauge
bosons ($W_H^+ \to W^+A_H \to e^+\nu_{e}A_H$ and $W_H^- \to W^-A_H
\to \mu^-\bar{\nu}_{\mu} A_H$). We take the branching ratio of the
$W_H$ boson decay as $Br(W_H \to W A_H)=100\%$ for $\kappa=1$ and
$f=800~GeV$ \cite{cpyuan:2006ph}, and the branching ratios of the
$W$ boson decay as $Br(W^+ \to e^+\nu_e)=10.75\%$ and $Br(W^- \to
\mu^-\bar{\nu}_\mu)= 10.57\%$ \cite{databook}. The $W_H$-pair
production channel including their subsequential decays can be
written as
\begin{eqnarray}\label{channel}
pp \to W_H^+ W_H^- \to W^+W^-A_HA_H \to
e^+\nu_{e}\mu^-\bar{\nu}_{\mu}A_H A_H.
\end{eqnarray}
Then a signal event of $W_H$-pair production is detected at the LHC
as two charged leptons ($e^+$ and $\mu^-$) plus missing energy
($A_HA_H\nu_{e}\bar{\nu}_\mu$). In Figs.\ref{fig8}(a) and (b) we
present the LO, QCD NLO corrected distributions of the transverse
momentum of $W^+$, and the corresponding $K$-factors at the
$\sqrt{s}=14~TeV$ and the $\sqrt{s}=8~TeV$ LHC, separately. From
Figs.\ref{fig8}(a,b) we can see that the QCD corrections always
enhance the LO differential cross section $d\sigma_{LO}/dp_T^{W^+}$,
and both the LO and QCD NLO corrected distributions of final $W^+$
boson at the future and the early LHC have their peaks around the
position of $p_T^{W^+}\sim 180~GeV$, and the $K$-factors are all
less than $1.20$.

\par
\begin{figure}[htbp]
\begin{center}
\includegraphics[width=0.45\textwidth]{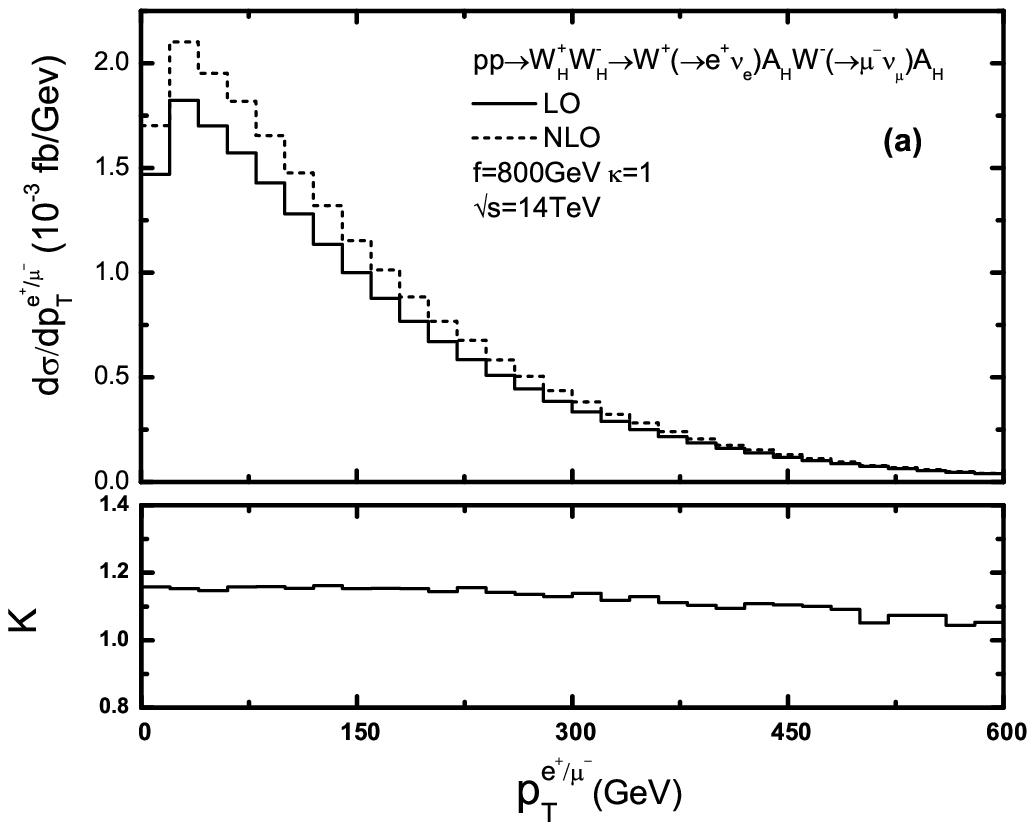}
\includegraphics[width=0.45\textwidth]{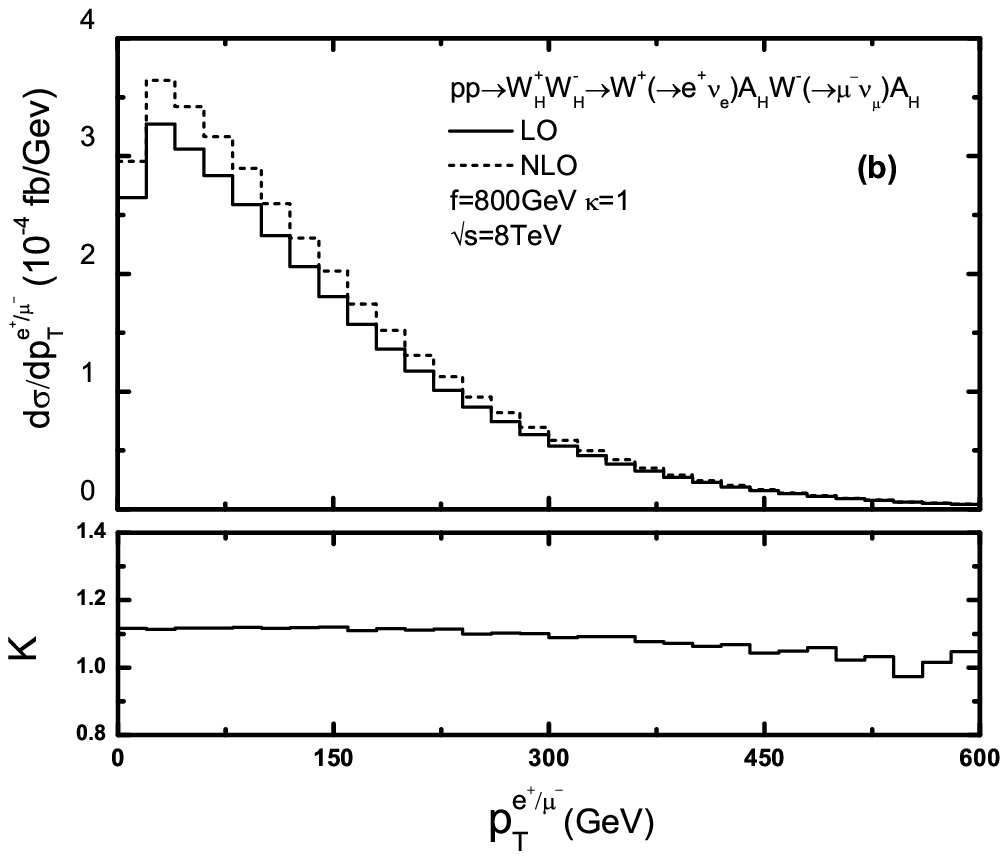}
\includegraphics[width=0.45\textwidth]{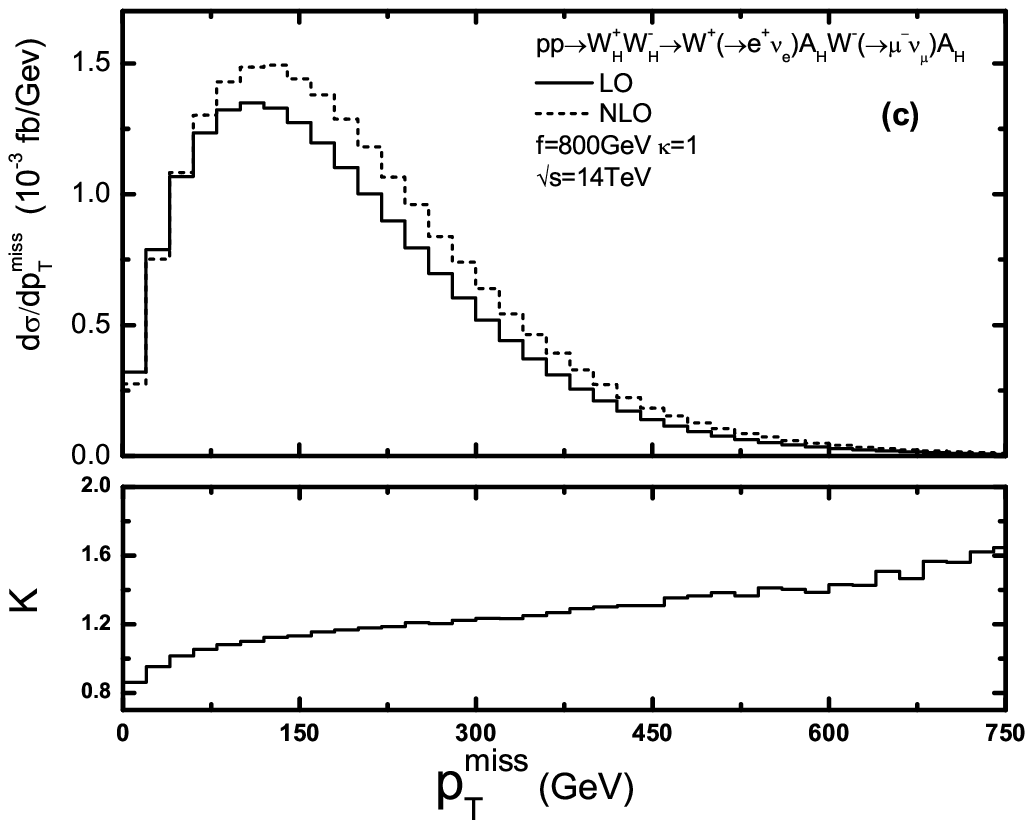}
\includegraphics[width=0.45\textwidth]{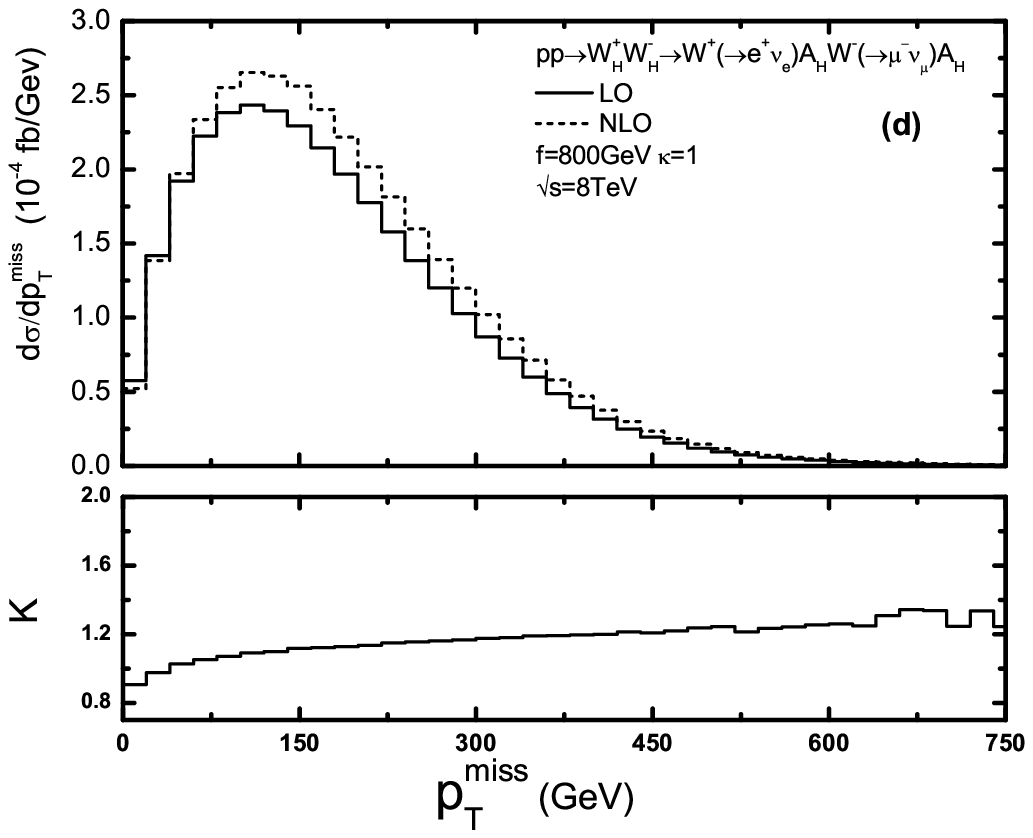}
\caption{\label{fig9} The LO, QCD NLO corrected transverse momentum
distributions of final particles and the corresponding $K$-factors of
the $pp \to W_H^+ W_H^- \to e^+ \mu^-  A_H A_H
\nu_{e}\bar{\nu}_\mu$ process at the LHC by taking $f=800~GeV$,
$\kappa=1$ and $s_\alpha = c_\alpha = \frac{\sqrt{2}}{2}$.
(a) $e^+/\mu^-$ transverse momentum distribution at the $\sqrt{s}=14~TeV$ LHC.
(b) $e^+/\mu^-$ transverse momentum distribution at the $\sqrt{s}=8~TeV$ LHC.
(c) missing transverse momentum distribution at the $\sqrt{s}=14~TeV$ LHC.
(d) missing transverse momentum distribution at the $\sqrt{s}=8~TeV$ LHC.}
\end{center}
\end{figure}

Since we neglect the masses of both positron and $\mu^-$ in
numerical calculations, the distribution of the positron transverse
momentum should be the same as $\mu^-$. The LO, QCD NLO corrected
distributions of the final lepton ($e^+/\mu^-$) transverse momentum
and missing transverse momentum of $A_H A_H \nu_{e}\bar{\nu}_\mu$,
and the corresponding $K$-factors are depicted in
Figs.\ref{fig9}(a,b,c,d) at the $\sqrt{s}=14~TeV$ and the
$\sqrt{s}=8~TeV$ LHC, separately. From Figs.\ref{fig9}(a,b) we can
see that the peaks on the $d\sigma_{LO}/dp_T^{e^+/\mu^-}$ and
$d\sigma_{NLO}/dp_T^{e^+/\mu^-}$ curves are all located at the
vicinity of $p_T \sim 30~GeV$. Figs.\ref{fig9}(c,d) show that the LO
and NLO missing transverse momentum distributions reach their maxima
at $p_T^{miss}\sim 110~GeV$, and the $K$-factors are between $0.86$
and $1.65$ at the $\sqrt{s}=14~TeV$ LHC, and between $0.91$ and
$1.24$ at the $\sqrt{s}=8~TeV$ LHC in the plotted range,
respectively. All the figures in Figs.\ref{fig8}(a,b) and
Figs.\ref{fig9}(a,b) show that the LO differential cross
sections, $\frac{d\sigma_{LO}}{dp_T^{W^{+}}}$,
$\frac{d\sigma_{LO}}{dp_T^{e^+/\mu^-}}$, at the early and the future LHC,
are enhanced by the QCD NLO calculations considerably in the
plotted $p_T$ regions, while the distributions of $p_T^{miss}$ in
Figs.\ref{fig9}(c,d) for the $14~TeV$ and the $8~TeV$ LHC,
show that the QCD NLO corrections decrease the LO differential
cross sections in the low $p_T^{miss}$ region.

\par
\begin{figure}[htbp]
\begin{center}
\includegraphics[width=0.45\textwidth]{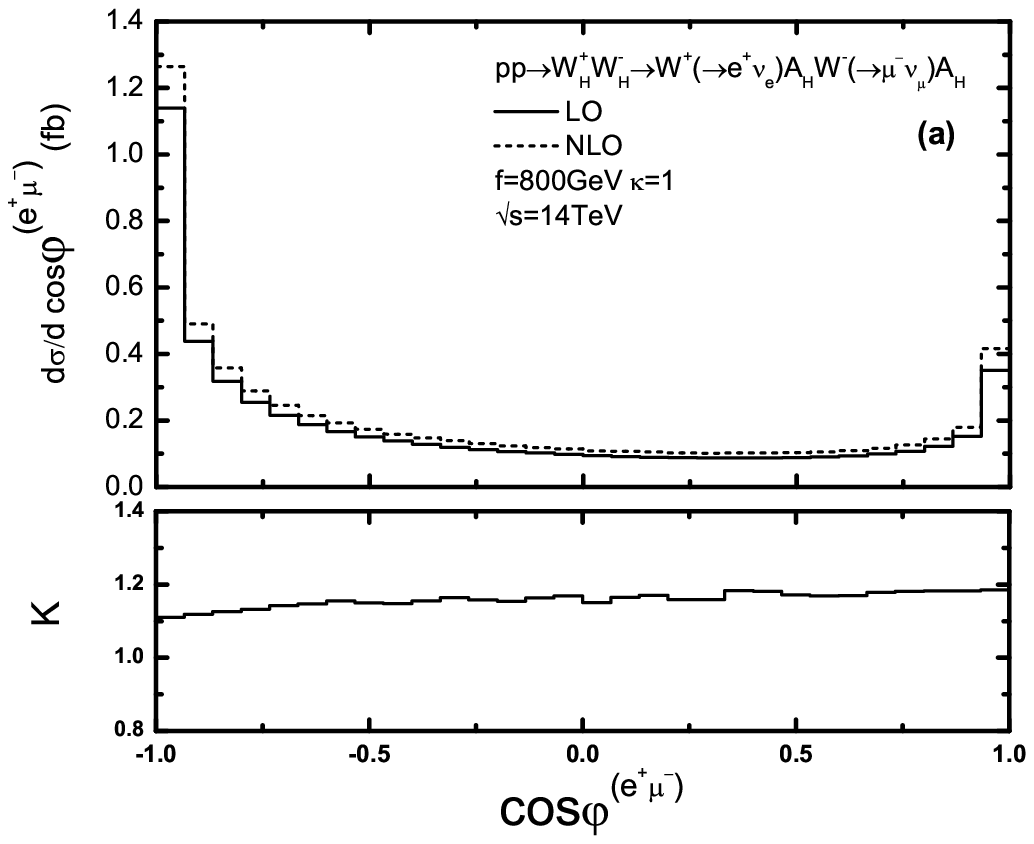}
\includegraphics[width=0.45\textwidth]{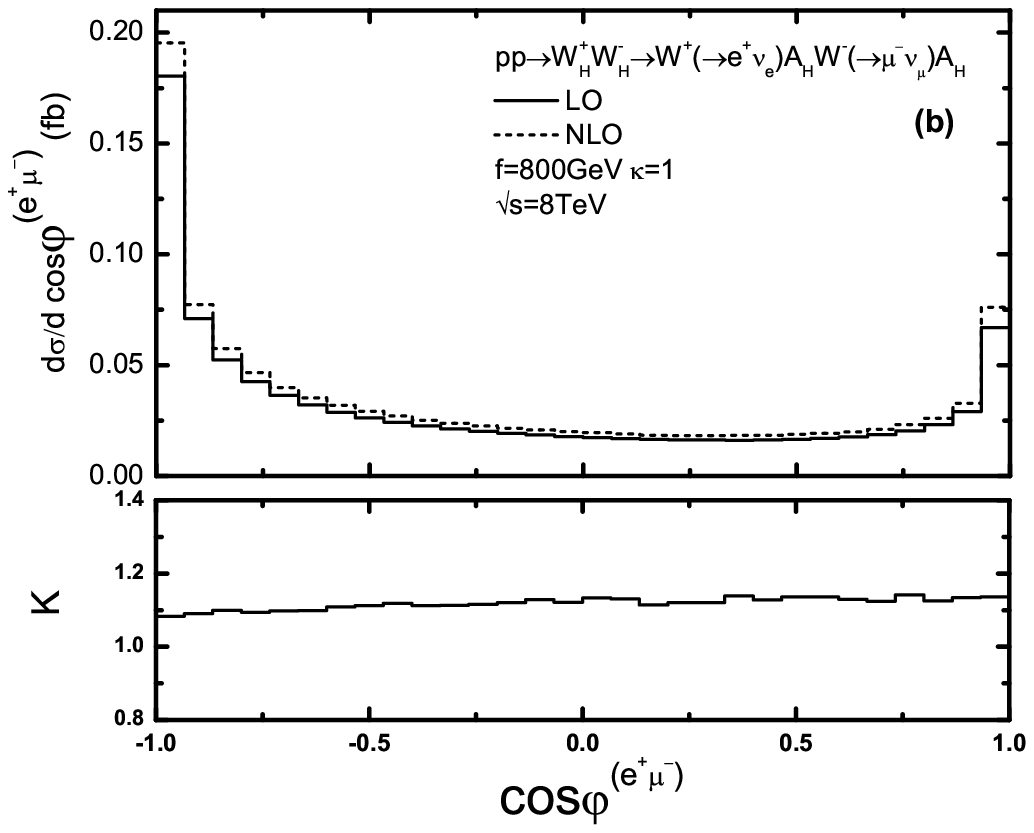}
\caption{\label{fig10a} The LO, QCD NLO distributions of $\cos
\varphi^{(e^+\mu^-)}$, where $\varphi^{(e^+\mu^-)}$ is the azimuthal
angle between leptons $e^+$ and $\mu^-$, and the corresponding
$K$-factors of the $pp \to W_H^+ W_H^- \to e^+ \mu^- A_H A_H
\nu_{e}\bar{\nu}_\mu$ process at the LHC by taking $f=800~GeV$,
$\kappa=1$ and $s_\alpha = c_\alpha = \frac{\sqrt{2}}{2}$. (a) $\sqrt{s}=14~TeV$.
(b) $\sqrt{s}=8~TeV$. }
\end{center}
\end{figure}

In Figs.\ref{fig10a}(a,b), we present the the LO, QCD NLO corrected
distributions of the azimuthal angle between $e^+$ and $\mu^-$ and
the corresponding $K$-factors for the $pp \to W_H^+ W_H^- \to e^+
\mu^- A_H A_H \nu_{e}\bar{\nu}_\mu$ process at the LHC, where we
take $f=800~GeV$, $\kappa=1$ and $s_\alpha = c_\alpha =
\frac{\sqrt{2}}{2}$ at the future $\sqrt{s}=14~TeV$ LHC and the
present $\sqrt{s}=8~TeV$ LHC, separately. The azimuthal
angle between $e^+$ and $\mu^-$ is obtained by using the following
equation:
\begin{eqnarray}
\varphi^{(e^+\mu^-)}=\arccos \left(
\frac{\vec{p}_T^{e^+}\cdot
\vec{p}_T^{\mu^-}}{|\vec{p}_T^{e^+}||\vec{p}_T^{\mu^-}|}
\right),
\end{eqnarray}
where $\vec{p}_T^{e^+}$ and $\vec{p}_T^{\mu^-}$ are the
three-momenta of $e^+$ and $\mu^-$. From the figures we can see that
the majority of the $W_H$-pair production events tends to have their
final leptons $e^+$ and $\mu^-$ outgoing back to back on the transverse plane.
That feature could explain the fact that the QCD NLO correction decreases the LO
differential cross section of $d\sigma_{LO}/dp_T^{miss}$ in the low
$p_T^{miss}$ region as shown in Figs.\ref{fig9}(c,d). As we know from
Fig.\ref{fig5}(a) that the correction part of $\Delta\sigma^{(2)}$
is usually negative, while the hard noncollinear real emission correction
part $\Delta\sigma^{(3)}$ is positive. The contribution of three-body term
from the real emission process $pp \to W_H^+W_H^-+{\rm jet} \to
e^+\mu^- + E_{\rm miss}+ {\rm jet}$ would kinematically suppress
$d \sigma/dp^{miss}_T$ in low $p_T^{miss}$ region comparing with the $p_T^{miss}$
distribution of the $pp \to W_H^+W_H^- \to e^+\mu^-+ E_{\rm miss}$ events
due to the majority of the final leptons $e^+$ and $\mu^-$ outgoing back
to back on the transverse plane.

\par
\begin{figure}[htbp]
\begin{center}
\includegraphics[width=0.45\textwidth]{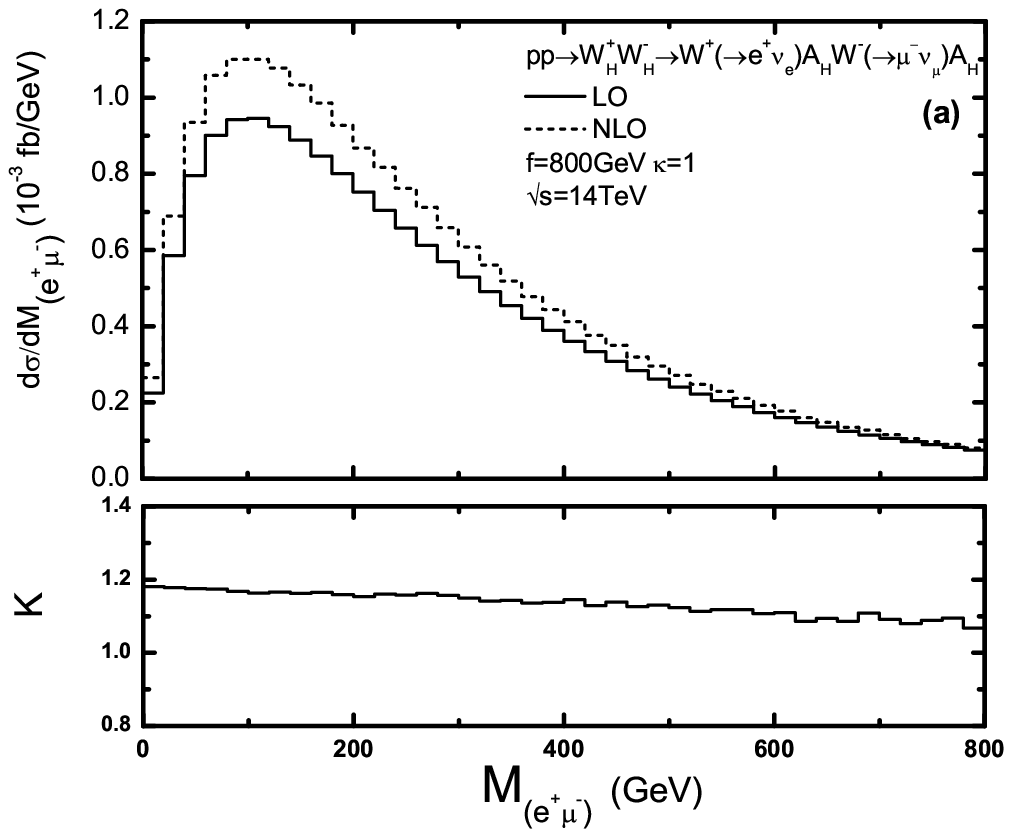}
\includegraphics[width=0.45\textwidth]{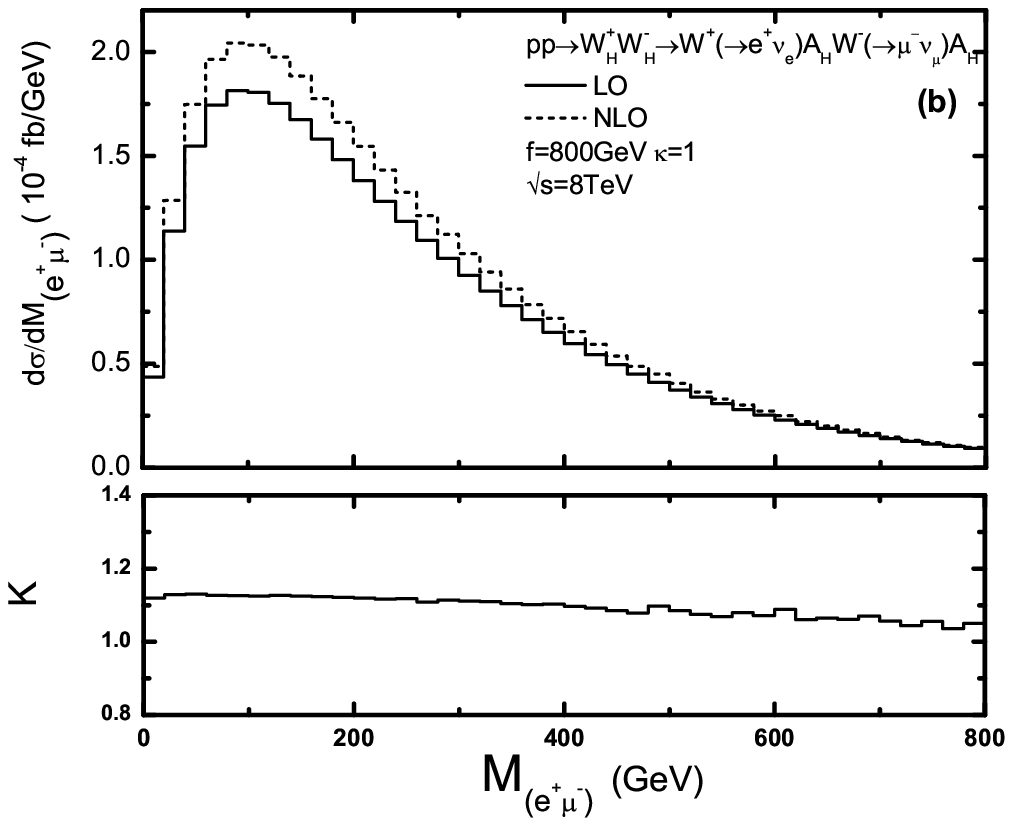}
\caption{\label{fig10b} The LO, QCD NLO distributions of invariant
mass of final positron and $\mu^-$, $M_{(e^+\mu^-)}$, and the
corresponding $K$-factors for the $pp \to W_H^+ W_H^- \to e^+ \mu^-
A_H A_H \nu_{e}\bar{\nu}_\mu$ process at the LHC by taking
$f=800~GeV$, $\kappa=1$ and $s_\alpha = c_\alpha =
\frac{\sqrt{2}}{2}$. (a) $\sqrt{s}=14~TeV$. (b) $\sqrt{s}=8~TeV$. }
\end{center}
\end{figure}

The LO and QCD NLO corrected distributions of the invariant mass of
final leptons, $e^+$ and $\mu^-$, are demonstrate in
Figs.\ref{fig10b}(a,b), where Figs.\ref{fig10b}(a) and (b) are for
the distributions of the invariant mass $M_{(e^+\mu^-)}$ and the
corresponding $K$-factors at the $\sqrt{s}=14~TeV$ and the
$\sqrt{s}=8~TeV$ LHC, respectively. We see from
Figs.\ref{fig10b}(a,b) that both LO and NLO curves for the
differential cross sections reach their maxima at the vicinity of
$M_{(e^+\mu^-)}\sim 100~GeV$.

\par
\begin{figure}[htbp]
\begin{center}
\includegraphics[width=0.45\textwidth]{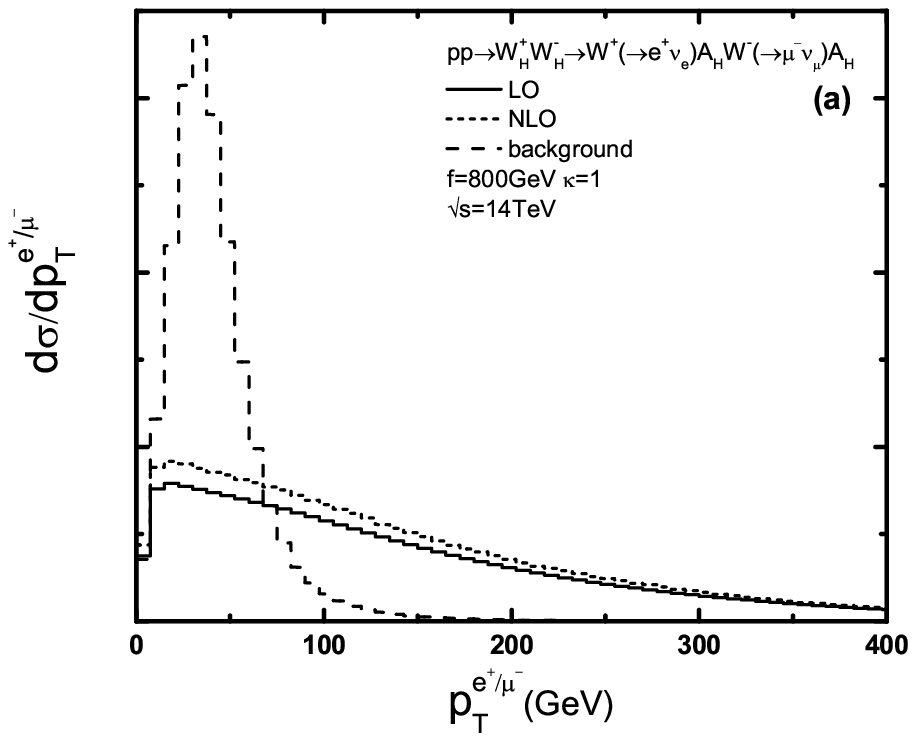}
\includegraphics[width=0.45\textwidth]{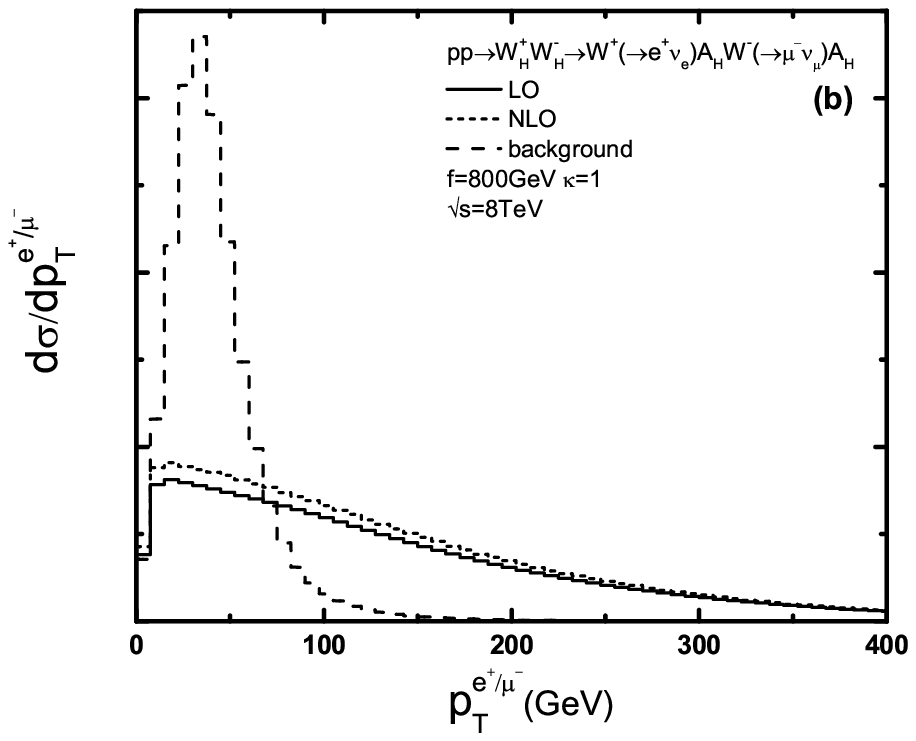}
\includegraphics[width=0.45\textwidth]{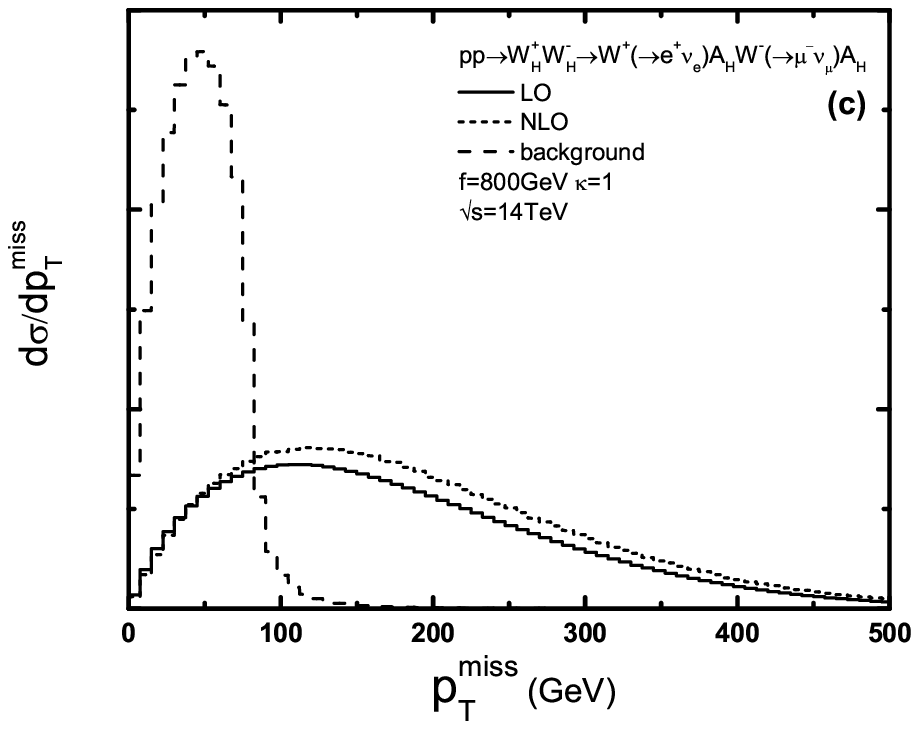}
\includegraphics[width=0.45\textwidth]{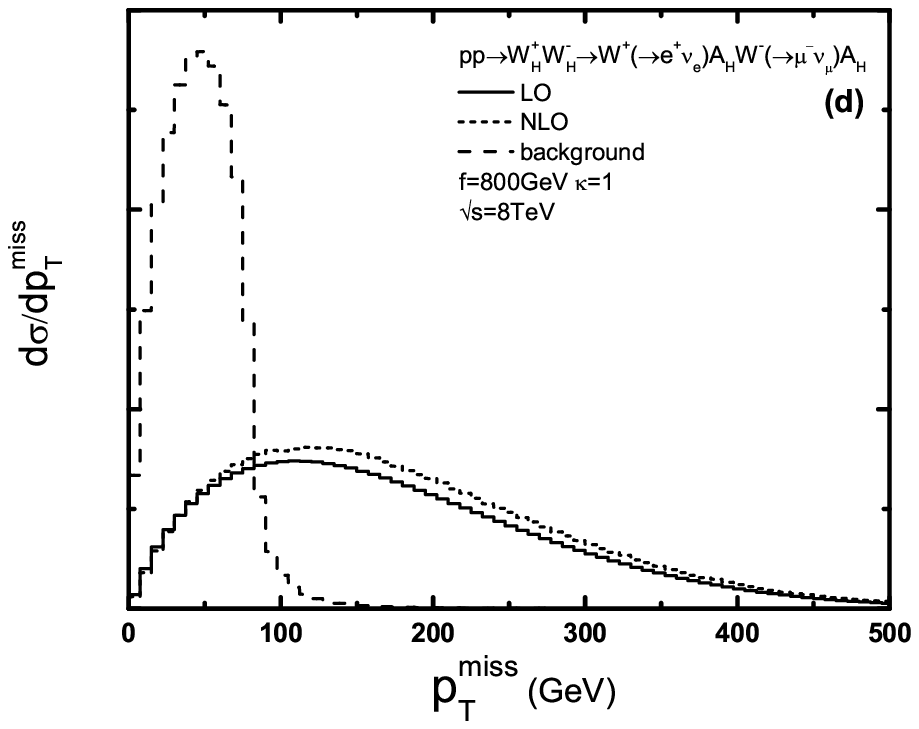}
\includegraphics[width=0.45\textwidth]{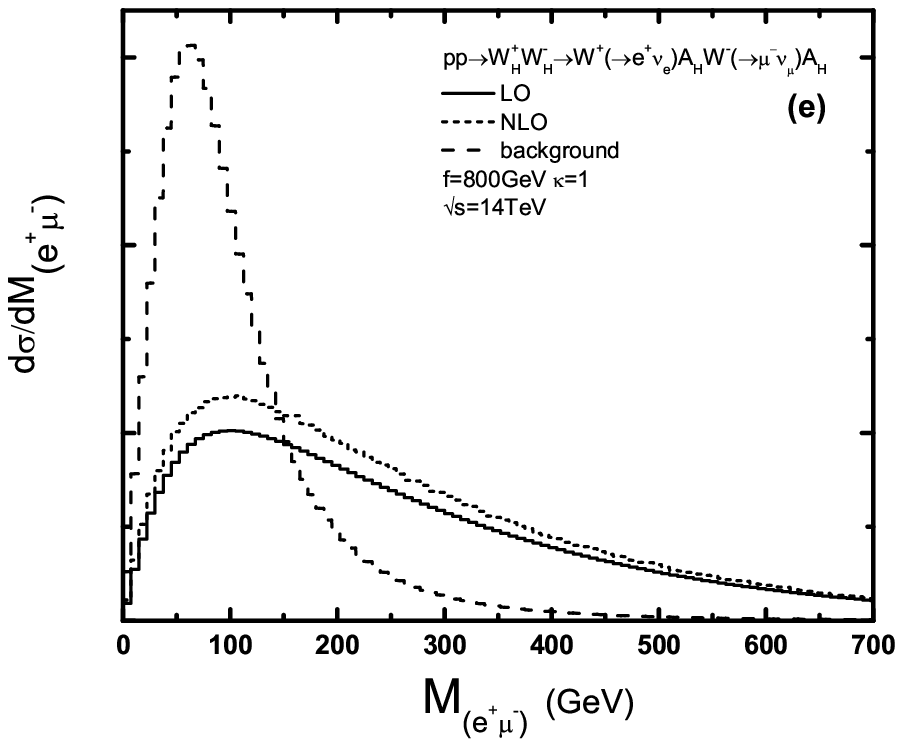}
\includegraphics[width=0.45\textwidth]{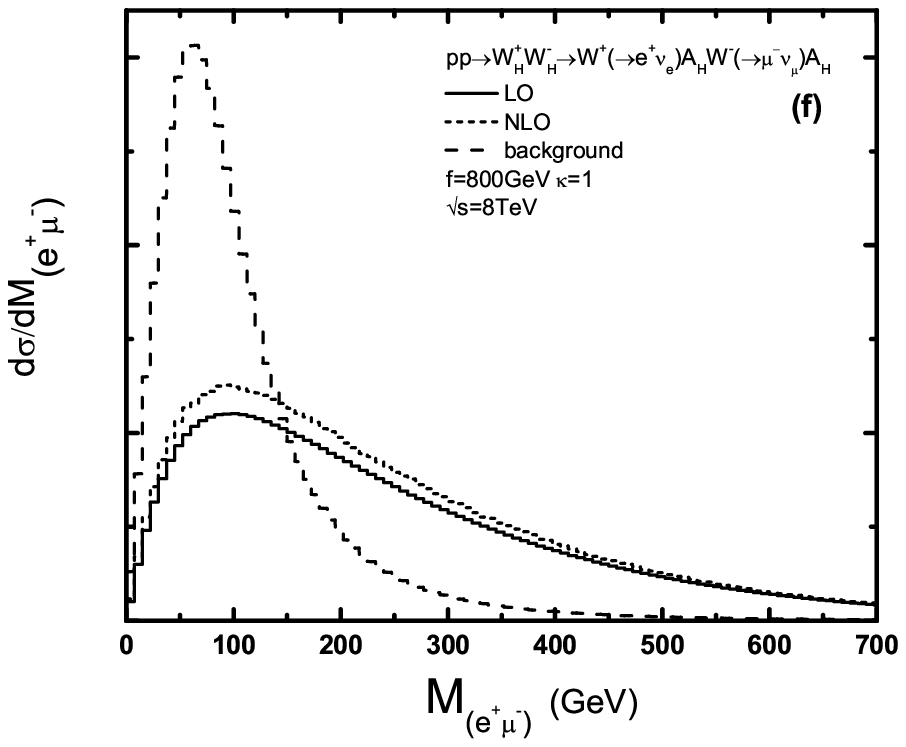}
\caption{\label{fig12} The LO, QCD NLO corrected distributions of
transverse momenta of final particles and the distribution of
invariant mass $M_{(e^+\mu^-)}$ of the signal process by taking
$f=800~GeV$, $\kappa=1$ and $s_\alpha = c_\alpha =
\frac{\sqrt{2}}{2}$. The background is from $pp \to W^+W^- \to
e^+\mu^- \nu_e\bar{\nu}_{\mu}$ process, and all curves in
Figs.\ref{fig12} are normalized by their total cross sections. (a)
$p_T$ distribution of final lepton $e^+/\mu^-$ at the
$\sqrt{s}=14~TeV$ LHC. (b) $p_T$ distribution of final lepton
$e^+/\mu^-$ at the $\sqrt{s}=8~TeV$ LHC. (c) missing transverse
momentum distribution at the $\sqrt{s}=14~TeV$ LHC. (d) missing
transverse momentum distribution at the $\sqrt{s}=8~TeV$ LHC. (e)
$M_{(e^+\mu^-)}$ distributions at the $\sqrt{s}=14~TeV$ LHC. (f)
$M_{(e^+\mu^-)}$ distributions at the $\sqrt{s}=8~TeV$ LHC. }
\end{center}
\end{figure}

As we know if the kinematic distribution of signal events is
distinctively different from that of background events, we can use
that feature to significantly suppress the background. For the
signal events of the $pp \to W_H^+ W_H^- \to e^+ \mu^- A_H A_H
\nu_{e}\bar{\nu}_\mu$ process, the dominant background at the LHC
comes from the SM process $pp\to W^+ W^- \to e^+ \mu^- \nu_{e}
\bar{\nu}_{\mu}+X$, which includes the same final leptons ($e^+$ and
$\mu^-$) and missing transverse momentum $p^{miss}_T$ due to final
$\nu_e\bar{\nu}_\mu$ products \cite{qhcao}. In order to show the
impact of the NLO corrections to the kinematic distributions, and
compare the distribution line shapes of signal and background, we
make the normalization procedure by dividing the differential cross
section by its LO total cross section. We plot the normalized
distributions of various kinematic observables of the signal of the
$W_H^+W_H^-$ pair production and its background in
Figs.\ref{fig12}(a-f) by taking $f=800~GeV$, $\kappa = 1$ and
$s_\alpha = c_\alpha = \frac{\sqrt{2}}{2}$. Figs.\ref{fig12}(a,c)
and Figs.\ref{fig12}(b,d) are for the $p_T$ distributions of final
lepton $e^+/\mu^-$ and undetectable particles at the
$\sqrt{s}=14~TeV$ and the $\sqrt{s}=8~TeV$ LHC, separately. In
Figs.\ref{fig12}(e,f) we show the normalized invariant mass
$M_{(e^+\mu^-)}$ distributions at the $\sqrt{s}=14~TeV$ and the
$\sqrt{s}=8~TeV$ LHC, respectively. We demonstrate the LO and QCD
NLO corrected distributions of $p_T^{e^+/\mu^-}$, $p_T^{miss}$ and
$M_{(e^+\mu^-)}$, together with the corresponding normalized LO
background distributions from $pp\to W^+ W^- \to e^+ \mu^- \nu_{e}
\bar{\nu}_{\mu}+X$ process. From those six figures, we can see that
the QCD NLO corrections obviously correct the LO differential cross
sections of final charged leptons, missing transverse momentum and
invariant mass $M_{(e^+\mu^-)}$ of the signal process, but do not
change their LO distribution line shapes very much. We see also that
the distributions of the background process $pp \to W^+W^- \to
e^+\mu^- \nu_e \bar{\nu}_{\mu}$ tend to be concentrated in the low
$p_T$ or low $M_{(e^+\mu^-)}$ range, while the signal distributions
can extend to more energetic ranges. Therefore, it is possible to
select the signal events of the $W_H$-pair production from its
background by taking proper $p_T$ or $M_{(e^+\mu^-)}$ lower
limits on the final charged leptons and the missing transverse
momentum. From Figs.\ref{fig12}(a-f) we can see if we take proper
lower limits on $p_T^{e^+/\mu^-}$, $p_T^{miss}$ and
$M_{(e^+\mu^-)}$, the background from $pp\to W^+ W^- \to e^+ \mu^-
\nu_{e}\bar{\nu}_{\mu}+X$ process can be significantly suppressed.

\vskip 5mm
\section{Summary}
\par
In this work, we present the calculation of the $W_H$-pair
production at the $\sqrt{s}=14~TeV$ and the $\sqrt{s}=8~TeV$ LHC in
QCD NLO. The dependence of the total cross section on the
renormalization/factorization scale shows that the QCD NLO
corrections reduce significantly the uncertainty of the LO
theoretical predictions from $48.88\%$ ($76.23\%$) to $13.40\%$
($14.54\%$) in the range of $\mu \in [0.1\mu_0, 10\mu_0]$ at the
$\sqrt{s}=14~TeV~(8~TeV)$ LHC. Our numerical results demonstrate
that the QCD NLO corrections enhance the LO integrated cross
sections with a $K$-factor in the range of $1.10-1.22$ ($1.09-1.17$)
as the global symmetry breaking scale $f$ varying from $400~GeV$
($400~GeV$) to $1.5~TeV$ ($1.0~TeV$) at the $\sqrt{s}=14~TeV$
$(8~TeV)$ LHC. We also investigate the kinematic distributions of
the transverse momenta of final $W$ boson, charged leptons and the
missing transverse momentum. We find that by putting proper lower
limits on $p_T$, $M_{(e^+\mu^-)}$ of the final leptons and the
missing transverse momentum it is possible to select the signal
events of the $W_H$-pair production from its background with high
ratio of signature to background.

\vskip 5mm
\par
\noindent{\large\bf Acknowledgments:} This work was supported in
part by the National Natural Science Foundation of China (Contract
No.11075150, No.11005101), and the Specialized Research Fund for the
Doctoral Program of Higher Education (Contract No.20093402110030).

\newpage
\section{Appendix }
\subsection{Appendix A: The relevant couplings}
\par
The Feynman rules for the coupling vertices in the LHT related to
our work are listed in Table \ref{tabA-1}
\cite{Hubisz:2004ft,THan,cpyuan:2006ph,KPan}, where we denote
$P_{L,R}=\frac{1}{2}(1\mp \gamma_5)$ and $v= v_{SM}$.
\begin{table}[h]
\tiny
\begin{center}
\begin{tabular}{|c|l||c|l|}
\hline
Vertex & ~~~~~~~~Feynman rule & Vertex & ~~~~~~~~~Feynman rule \\
\hline
&&& \\
$W_{H}^{+\mu}(k_1)W_{H}^{-\nu}(k_2)A^\rho(k_3)$ & $-ie \left[ g^{\mu
\nu} (k_1-k_2)^\rho+\right.$ &
$W_{H}^{+\nu}(k_1)W_{H}^{-\nu}(k_2)Z^\rho(k_3)$ &
$-ie\frac{c_w}{s_w}
\left[g^{\mu \nu} (k_1-k_2)^\rho+\right.$\\
&&& \\
 & $\left. g^{\nu \rho} (k_2-k_3)^\mu+g^{\rho \mu} (k_3-k_1)^\nu \right]$ &
 & $\left. +g^{\nu \rho}(k_2-k_3)^\mu+g^{\rho \mu} (k_3-k_1)^\nu \right]$\\
&&& \\
$W^+_{H\mu} \bar U_j D_{i-}(i=1,2,j=1,2,3)$ & $i\frac{g}{\sqrt{2}}
\gamma_\mu P_L(V_{Hu})_{ij}$ &
$W^-_{H\mu} \bar D_j U_{i-}(i,j=1,2,3)$ & $i\frac{g}{\sqrt{2}} \gamma_\mu P_L(V_{Hd})_{ij}$ \\
&&& \\
$W^+_{H\mu} \bar t D_{j-}(j=1,2,3)$ & $i\frac{g}{\sqrt{2}}
\gamma_\mu c_L P_L(V_{Hu})_{j3} $ &
$h W^+_{H\mu} W^-_{H\nu}$ & $-ie\frac{m_W}{s_w} g_{\mu \nu}$ \\
&&& \\
$W^-_{H\mu} \bar T_+ D_{j-}(j=1,2,3)$ & $i\frac{g}{\sqrt{2}}
\gamma_\mu s_L P_L (V_{Hu})_{j3}$ &
$h \bar T_+ T_+$ & $i\frac{m_t c_\alpha s_\alpha}{f}$ \\
&&& \\
$h \bar U_{i-} U_{i-}(i=1,2,3)$ & $i\frac{\sqrt{2}\kappa v}{4f}$ &
$h \bar t t$ &
$-i\frac{m_t}{v}\left[1-\left(\frac{3}{4}-c_\alpha^2+c_\alpha^4\right)
\frac{v^2}{f^2}\right]$ \\
&&& \\
$Z_\mu \bar U_{i-} U_{i-}(i=1,2,3)$ & $i\frac{g}{c_w}
    \gamma_\mu\left(\frac{1}{2}-\frac{2}{3}s_w^2\right)$ & $Z_\mu \bar T_- T_-$ &
$i\frac{g}{c_w} \gamma_\mu\left(-\frac{2}{3}s_w^2\right)$ \\
&&& \\
$Z_\mu \bar D_{i-} D_{i-}(i=1,2,3)$ & $i\frac{g}{c_w}
    \gamma_\mu\left(-\frac{1}{2}+\frac{1}{3}s_w^2\right)$ &
$$ & $$ \\ &&& \\
$Z_\mu \bar t t$ & $i\frac{g}{c_w} \gamma_\mu \left[
    \left(\frac{1}{2}-\frac{2}{3}s_w^2-\frac{s_\alpha^4}{2}\frac{v^2}{f^2}\right)P_L\right.$
    & $Z_\mu \bar T_+ T_+$ & $i\frac{g}{c_w} \gamma_\mu \left[
    \left(-\frac{2}{3}s_w^2+\frac{s_\alpha^4}{2}\frac{v^2}{f^2}\right)P_L \right. $ \\
&&& \\
 & $\left. ~~~~~~~~ -\frac{2}{3} s_w^2 P_R \right]$ &
 & $\left.~~~~~~~~  -\frac{2}{3} s_w^2 P_R \right]$ \\
&&& \\
$Z_{H\mu}\bar U_j U_{i-}(i,j=1,2)$ & $i\left(\frac{g
c_H}{2}-\frac{g' s_H}{10}\right)\gamma_\mu P_L (V_{Hu})_{ij}$ &
$A_{H\mu}\bar U_j U_{i-}(i,j=1,2)$ & $i\left(-\frac{g
s_H}{2}-\frac{g'
c_H}{10}\right)\gamma_\mu P_L (V_{Hu})_{ij}$\\
&&& \\
&&& \\
$Z_{H\mu}\bar D_j D_{i-}(i,j=1,2,3)$ & $i\left(-\frac{g
c_H}{2}-\frac{g' s_H}{10}\right)\gamma_\mu P_L (V_{Hd})_{ij}$ &
$A_{H\mu}\bar D_j D_{i-}(i,j=1,2,3)$ & $i\left(\frac{g
s_H}{2}-\frac{g' c_H}{10}\right)\gamma_\mu P_L (V_{Hd})_{ij}$\\
&&&\\
&&& \\
$Z_{H\mu}\bar t t_-$ & $i\left(\frac{g c_H}{2}-\frac{g'
s_H}{10}\right)c_L\gamma_\mu P_L(V_{Hu})_{33}$ &
$A_{H\mu}\bar t t_-$ & $i\left(-\frac{g s_H}{2}-\frac{g' c_H}{10}\right)c_L\gamma_\mu P_L(V_{Hu})_{33} $\\
&&& \\

$G_\mu^a \bar q_-^\alpha q_-^\beta$ & $ig_s(T^a)_{\alpha \beta}
\gamma_\mu$ & $G_\mu^a \bar T_{\pm}^\alpha T_{\pm}^\beta$ &
$ig_s(T^a)_{\alpha \beta} \gamma_\mu$ \\
&&& \\
\hline
\end{tabular}
\caption{\label{tabA-1} The related LHT Feynman rules used in our
calculations. There $c_L=\sqrt{1-s_\alpha^4\frac{v^2}{f^2}}$,
$s_L=s_\alpha^2\frac{v}{f}$,  $U_i=u,c,t$, $D_i=d,s,b$,
$U_{i-}=u_-,c_-,t_-$ and $D_{i-}=d_-,s_-,b_-$. $i$ and $j$ are the
generation indices. }
\end{center}
\end{table}

\par
\subsection{Appendix B: Partial decay widths}
\par
The partial decay widths of $T$-odd up-type and down-type quarks can be
generally expressed as
\begin{eqnarray} \label{Width-1}
\Gamma (U_{i-}\to W_H^+ D_j) &=&
\frac{g^2|(V_{Hd})_{ij}|^2}{64\pi}\frac{m_{U_{i-}}^3}{m_{W_H}^2}
\left[\left(1- \frac{m_{W_H}^2}{m_{U_{i-}}^2}\right) \left(1+
\frac{2m_{W_H}^2}{m_{U_{i-}}^2}\right)+ \right.
\nb \\
&+&\left.\frac{m_{D_j}^2}{m_{U_{i-}}^4}\left(m_{D_j}^2+m_{W_H}^2-2m_{U_{i-}}^2\right)\right]
\left[\left(1-\left(\frac{m_{W_H}+m_{D_j}}{m_{U_{i-}}}\right)^2\right)
\left(1-\left(\frac{m_{W_H}-m_{D_j}}{m_{U_{i-}}}\right)^2\right)\right]^{\frac{1}{2}}, \nb \\
&&~~~~~~~~~~~~~~~~~~~~~~~~~~~~~~~~~~~~~~~~~~~~~~~~~~~~~~~~~~~~~~~~~~~~~~~~~~~~~~~~~~(i,j=1,2,3),\nb \\
\Gamma (U_{i-}\to Z_H U_j) &=& \frac{2|(V_{Hu})_{ij}|^2\left(\frac{g
c_H}{2} - \frac{g' s_H}{10}\right)^2}{64\pi}\frac{m_{U_{i-}}^3}
{m_{Z_H}^2} \left[\left(1- \frac{m_{Z_H}^2}{m_{U_{i-}}^2}\right) \left(1+
\frac{2m_{Z_H}^2}{m_{U_{i-}}^2}\right)+ \right.\nb \\
&+&\left.\frac{m_{U_j}^2}{m_{U_{i-}}^4}\left(m_{U_j}^2+m_{Z_H}^2-2m_{U_{i-}}^2\right)\right]
\left[\left(1-\left(\frac{m_{Z_H}+m_{U_j}}{m_{U_{i-}}}\right)^2\right)
\left(1-\left(\frac{m_{Z_H}-m_{U_j}}{m_{U_{i-}}}\right)^2\right)\right]^{\frac{1}{2}},\nb \\
&&~~~~~~~~~~~~~~~~~~~~~~~~~~~~~~~~~~~~~~~~~~~~~~~~~~~~~~~~~~~~~~~~~~~~~~~~~~~~~~~~~~(i,j=1,2), \nb \\
\Gamma (U_{i-}\to A_H U_j) &=& \frac{2|(V_{Hu})_{ij}|^2\left(\frac{g
s_H}{2} + \frac{g' c_H}{10}\right)^2}{64\pi}\frac{m_{U_{i-}}^3}
{m_{A_H}^2} \left[\left(1- \frac{m_{A_H}^2}{m_{U_{i-}}^2}\right)
\left(1+ \frac{2m_{A_H}^2}{m_{U_{i-}}^2}\right)+ \right.
\nb \\
&+&\left.\frac{m_{U_j}^2}{m_{U_{i-}}^4}\left(m_{U_j}^2+m_{A_H}^2-2m_{U_{i-}}^2\right)\right]
\left[\left(1-\left(\frac{m_{A_H}+m_{U_j}}{m_{U_{i-}}}\right)^2\right)
\left(1-\left(\frac{m_{A_H}-m_{U_j}}{m_{U_{i-}}}\right)^2\right)\right]^{\frac{1}{2}}, \nb \\
&&~~~~~~~~~~~~~~~~~~~~~~~~~~~~~~~~~~~~~~~~~~~~~~~~~~~~~~~~~~~~~~~~~~~~~~~~~~~~~~~~~~(i,j=1,2),\nb \\
\Gamma (t_{-}\to Z_H t) &=&
\frac{2|(V_{Hu})_{33}|^2\left(\frac{gc_H}{2} - \frac{g'
s_H}{10}\right)^2}{64\pi}\frac{c_L^2 m_{t_{-}}^3} {m_{Z_H}^2}
\left[\left(1- \frac{m_{Z_H}^2}{m_{t_{-}}^2}\right) \left(1+
\frac{2m_{Z_H}^2}{m_{t_{-}}^2}\right)+ \right.
\nb \\
&+&\left.\frac{m_{t}^2}{m_{t_{-}}^4}\left(m_{t}^2+m_{Z_H}^2-2m_{t_{-}}^2\right)\right]
\left[\left(1-\left(\frac{m_{Z_H}+m_{t}}{m_{t_{-}}}\right)^2\right)
\left(1-\left(\frac{m_{Z_H}-m_{t}}{m_{t_{-}}}\right)^2\right)\right]^{\frac{1}{2}},
\nb \\
\Gamma (t_{-}\to A_H t) &=& \frac{2|(V_{Hu})_{33}|^2\left(\frac{g
s_H}{2} + \frac{g' c_H}{10}\right)^2}{64\pi}\frac{c_L^2m_{t_{-}}^3}
{m_{A_H}^2} \left[\left(1- \frac{m_{A_H}^2}{m_{t_{-}}^2}\right)
\left(1+ \frac{2m_{A_H}^2}{m_{t_{-}}^2}\right)+ \right.
\nb \\
&+&\left.\frac{m_{t}^2}{m_{t_{-}}^4}\left(m_{t}^2+m_{A_H}^2-2m_{t_{-}}^2\right)\right]
\left[\left(1-\left(\frac{m_{A_H}+m_{t}}{m_{t_{-}}}\right)^2\right)
\left(1-\left(\frac{m_{A_H}-m_{t}}{m_{t_{-}}}\right)^2\right)\right]^{\frac{1}{2}}, \nb \\
\end{eqnarray}

\begin{eqnarray}\label{Width-2}
\Gamma (D_{i-}\to W_H^- U_j) &=&
\frac{g^2|(V_{Hu})_{ij}|^2}{64\pi}\frac{m_{D_{i-}}^3}{m_{W_H}^2}
\left[\left(1- \frac{m_{W_H}^2}{m_{D_{i-}}^2}\right) \left(1+
\frac{2m_{W_H}^2}{m_{D_{i-}}^2}\right)+ \right.
\nb \\
&+&\left.\frac{m_{U_j}^2}{m_{D_{i-}}^4}\left(m_{U_j}^2+m_{W_H}^2-2m_{D_{i-}}^2\right)\right]
\left[\left(1-\left(\frac{m_{W_H}+m_{U_j}}{m_{D_{i-}}}\right)^2\right)
\left(1-\left(\frac{m_{W_H}-m_{U_j}}{m_{D_{i-}}}\right)^2\right)\right]^{\frac{1}{2}}, \nb\\
&&~~~~~~~~~~~~~~~~~~~~~~~~~~~~~~~~~~~~~~~~~~~~~~~~~~~~~~~~~~~~~~~~~~~~~~~~~~(i=1,2, ~j=1,2,3),\nb \\
\Gamma (b_{-}\to W_H^- t) &=&
\frac{g^2|(V_{Hu})_{33}|^2}{64\pi}\frac{c_L^2m_{b_{-}}^3}{m_{W_H}^2}
\left[\left(1- \frac{m_{W_H}^2}{m_{b_{-}}^2}\right) \left(1+
\frac{2m_{W_H}^2}{m_{b_{-}}^2}\right)+ \right.
\nb \\
&+&\left.\frac{m_{t}^2}{m_{b_{-}}^4}\left(m_{t}^2+m_{W_H}^2-2m_{b_{-}}^2\right)\right]
\left[\left(1-\left(\frac{m_{W_H}+m_{t}}{m_{b_{-}}}\right)^2\right)
\left(1-\left(\frac{m_{W_H}-m_{t}}{m_{b_{-}}}\right)^2\right)\right]^{\frac{1}{2}}, \nb\\
\Gamma (D_{i-}\to Z_H D_{j}) &=& \frac{2|(V_{Hd})_{ij}|^2
\left(\frac{g c_H}{2} + \frac{g's_H}{10}\right)^2}{64\pi}\frac{m_{D_{i-}}^3}
{m_{Z_H}^2}\left[\left(1- \frac{m_{Z_H}^2}{m_{D_{i-}}^2}\right) \left(1+
\frac{2m_{Z_H}^2}{m_{D_{i-}}^2}\right)+ \right.
\nb \\
&+&\left.\frac{m_{D_j}^2}{m_{D_{i-}}^4}\left(m_{D_j}^2+m_{Z_H}^2-2m_{D_{i-}}^2\right)\right]
\left[\left(1-\left(\frac{m_{Z_H}+m_{D_j}}{m_{D_{i-}}}\right)^2\right)
\left(1-\left(\frac{m_{Z_H}-m_{D_j}}{m_{D_{i-}}}\right)^2\right)\right]^{\frac{1}{2}},
\nb \\
&&~~~~~~~~~~~~~~~~~~~~~~~~~~~~~~~~~~~~~~~~~~~~~~~~~~~~~~~~~~~~~~~~~~~~~~~~~~~~~~~~~~(i,j=1,2,3),\nb \\
\Gamma (D_{i-}\to A_H D_j) &=& \frac{2|(V_{Hd})_{ij}|^2\left(\frac{g
s_H}{2} - \frac{g' c_H}{10}\right)^2}{64\pi}\frac{m_{D_{i-}}^3}
{m_{A_H}^2}\left[\left(1- \frac{m_{A_H}^2}{m_{D_{i-}}^2}\right) \left(1+
\frac{2m_{A_H}^2}{m_{D_{i-}}^2}\right)+ \right.
\nb \\
&+&\left.\frac{m_{D_j}^2}{m_{D_{i-}}^4}\left(m_{D_j}^2+m_{A_H}^2-2m_{D_{i-}}^2\right)\right]
\left[\left(1-\left(\frac{m_{A_H}+m_{D_j}}{m_{D_{i-}}}\right)^2\right)
\left(1-\left(\frac{m_{A_H}-m_{D_j}}{m_{D_{i-}}}\right)^2\right)\right]^{\frac{1}{2}}. \nb \\
&&~~~~~~~~~~~~~~~~~~~~~~~~~~~~~~~~~~~~~~~~~~~~~~~~~~~~~~~~~~~~~~~~~~~~~~~~~~~~~~~~~~(i,j=1,2,3),\nb \\
\end{eqnarray}
The LO total decay width of the $T$-odd quark $q_-$ can be obtained
approximately by summing up all the LO partial decay widths of the
main decay channels. In Eqs.(\ref{Width-1}) and (\ref{Width-2}),
$U_{i-}=u_-,c_-,t_-$, $D_{i-}=d_-,s_-,b_-$, $U_i=u,c,t$ and
$D_i=d,s,b$.

\end{document}